\documentclass[10pt,twocolumn]{IEEEtran}
\usepackage{algorithm}
\usepackage{algorithmic}
\usepackage{graphicx}
\usepackage{amsmath,amsthm,amssymb}
\usepackage{subfigure}
\usepackage[utf8]{inputenc}
\usepackage{float}
\usepackage{cite}
\usepackage{flushend}

\newtheorem{theorem}{Theorem}
\newtheorem{definition}{Definition}

\floatstyle{ruled}
\newfloat{procedure}{t}{lop}
\floatname{procedure}{Procedure}
\newfloat{algo}{t}{lop}
\floatname{algo}{Algorithm}

\newcommand{\la}{\leftarrow}

\begin{document}

\title{Adaptively Sharing Real-Time Aggregate\\ with Differential Privacy}

\author{Liyue~Fan,
        Li~Xiong
\IEEEcompsocitemizethanks{\IEEEcompsocthanksitem L. Fan and L. Xiong are with the Department
of Mathematics and Computer Science, Emory University, Atlanta,
GA, 30322.\protect\\}
\thanks{}}

\IEEEcompsoctitleabstractindextext{%
\begin{abstract}
Sharing real-time aggregate statistics of private data is of great value to the public to perform data mining for understanding important phenomena, such as Influenza outbreaks and traffic congestion. However, releasing time-series data with standard differential privacy mechanism has limited utility due to high correlation between data values. We propose FAST, a novel framework to release real-time aggregate statistics under differential privacy based on filtering and adaptive sampling.  
To minimize the overall privacy cost,  FAST adaptively samples long time-series according to the detected data dynamics.  To improve the accuracy of data release per time stamp, FAST predicts data values at non-sampling points and corrects noisy observations at sampling points.  Our experiments with real-world as well as synthetic data sets confirm that FAST improves the accuracy of released aggregates even under small privacy cost and can be used to enable a wide range of monitoring applications. 
\end{abstract}

}

\maketitle

\IEEEdisplaynotcompsoctitleabstractindextext

\IEEEpeerreviewmaketitle

\section{Introduction}

\IEEEPARstart{S}{haring} real-time aggregate statistics of private data enables many important data mining applications.   Consider the examples below:
\vspace{0.1cm}
\begin{itemize}
\item[]
\textbf{Disease Surveillance}: A health care provider gathers data from individual visitors.  The collected data, e.g. daily number of Influenza cases,  is then shared with third parties, e.g., researchers, in order to monitor and to detect seasonal epidemic outbreaks at the earliest. 
\item[]
\textbf{Traffic Monitoring}: A GPS service provider gathers data from individual users about their locations, speeds, mobility,  etc.  The aggregated data, for instance, the number of users at each region during each time period, can be mined for commercial interest, such as popular places, as well as public interest, such as congestion patterns on the roads.
\end{itemize}

\begin{figure}
\centering
\includegraphics[width=0.8\columnwidth]{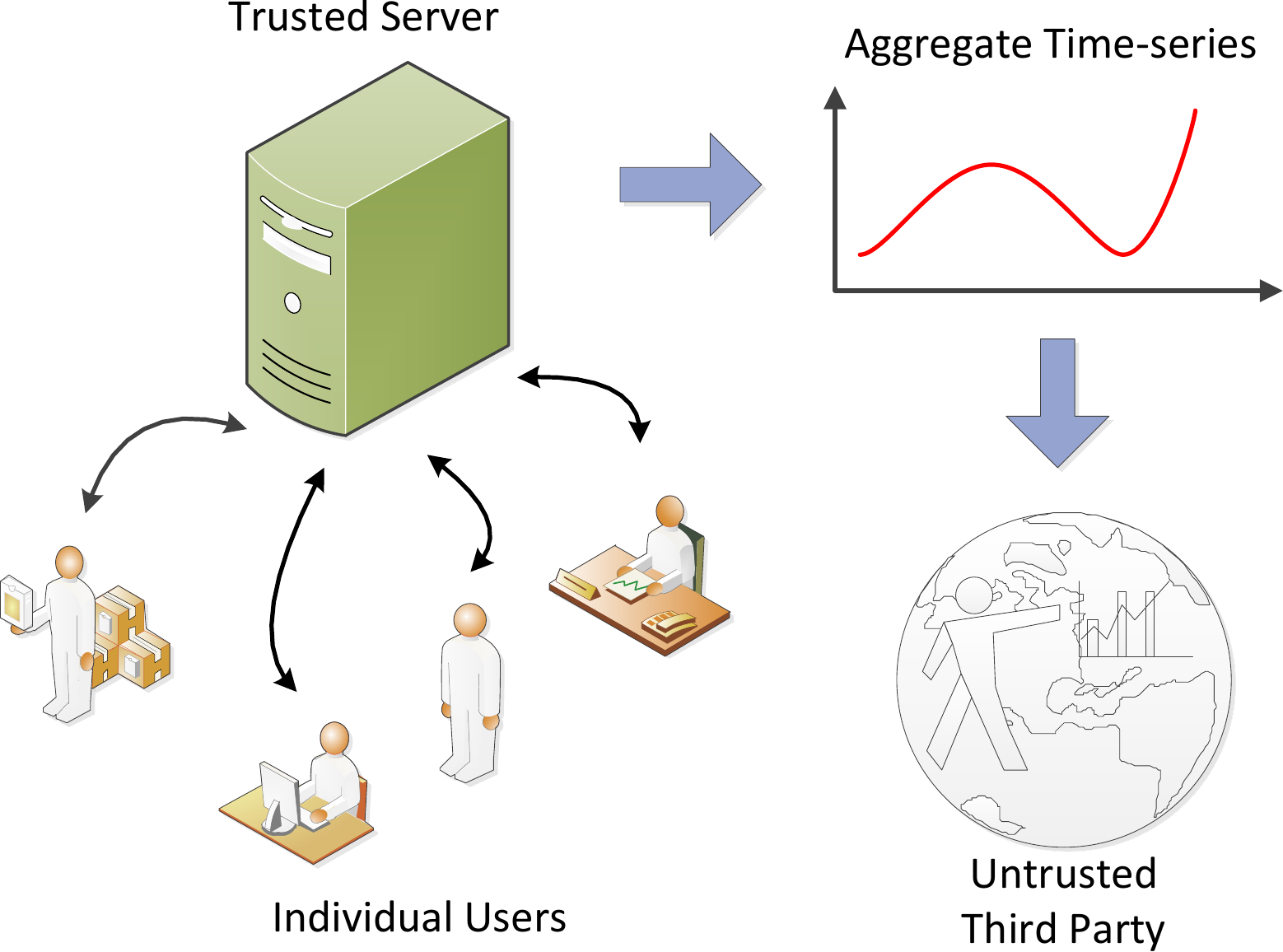}
\caption{\small Aggregate Data Sharing Scenario}
\label{fig:sharing}
\vspace{-0.5cm}
\end{figure}

A common scenario of such applications can be summarized by Figure~\ref{fig:sharing}, where a trusted server gathers data from a large number of individual subscribers. The collected data may be then aggregated and continuously shared with other un-trusted entities for various purposes. The trusted server, i.e. publisher, is assumed to be bound by contractual obligations to protect the user's interests, therefore it must ensure that releasing the data does not compromise the privacy of any individual who contributed data.   The goal of our work is to enable the publisher to share useful aggregate statistics over individual users continuously (aggregate time series) while guaranteeing their privacy.

The current state-of-the-art paradigm for privacy-preserving data publishing is \textit{differential privacy}\cite{BLR08}, which requires that the aggregate statistics reported by a data publisher be perturbed by a randomized algorithm $\mathcal{A}$, so that the output of $\mathcal{A}$ remains roughly the same even if any single tuple in the input data is arbitrarily modified.  Given the output of $\mathcal{A}$, an adversary will not be able to infer much about any single tuple in the input, and thus privacy is protected.

Most existing works on differentially private data release deal with one-time release of static data \cite{Dwork06, Hay09, Xiao11, Xiao11TKDE, Xiao10, Xu12}.   In the applications we consider, 
  data values at successive timestamps are highly correlated. A straightforward application of differential privacy mechanism which adds a Laplace noise to each aggregate value at each time stamp can lead to a very high overall perturbation error due to the composition theorem~\cite{McSherry09}.
Few recent works \cite{ChanSS10, Dwork10, RN10} studied the problem of releasing time series or continual statistics.  Rastogi and Nath~\cite{RN10} proposed an algorithm which perturbs $d$ Discrete Fourier Transform (DFT) coefficients of the entire time series and reconstructs a released version with the Inverse DFT series.  Since the entire time-series is required to perform those operations, it is not applicable to real-time applications.
Dwork et al.~\cite{Dwork10} proposed a differentially private continual counter over a binary stream with a bounded error at each time step.
Chan et al.~\cite{ChanSS10} studied the same problem and concluded with a similar upper bound.  However, both works adopt an event-level privacy model, with the perturbation mechanism designed to protect the presence of an individual event, i.e. a user's contribution to the data stream at a single time point, rather than the presence or privacy of a user.

In this paper, we present FAST, a real-time system with \underline{F}iltering and \underline{A}daptive \underline{S}ampling for differentially private \underline{T}ime-series monitoring. It uses sampling to query and perturb selected values in the time series with the differential privacy mechanism,  and simultaneously uses filtering to dynamically predict the non-sampled values and correct the sampled values.  The key innovation is that FAST utilizes feedback loops based on observed (perturbed) values to dynamically adjust the prediction/estimation model as well as the sampling rate. To this end, we examine two challenges in our system: predictability and controllability. The former raises the question: given a perturbed value, can we formulate an estimate
which is close to the true value and dynamically adjust the estimation model based on current observation? The latter imposes another question: suppose an accurate estimate can be derived at any time step, can we dynamically adjust the sampling rate according to the rate of data change?  We extend our recent work~\cite{Fan12CIKM} and present several contributions: 

1) To improve the accuracy of data release at each time stamp, we establish the state space model for time-series data and use filtering techniques to estimate the original data values from observed values.    By assuming a process model that characterizes the time series, we can reduce the impact of perturbation errors introduced by the differential privacy mechanism.   This is achieved by combining the noisy observation with a prediction based on the process model.  The combined value, also referred to as correction or posterior estimate, provides an educated guess rather than a pure perturbed answer.  The posterior estimate is then fed back to the system for future predictions and for dynamically adjusting the sampling process.

2) To minimize the overall privacy cost, hence, the overall perturbation error, we propose to sample the time series data as needed.   Besides the fixed-rate sampling method, we introduce an adaptive sampling algorithm which adjusts the sampling rate with PID  control (stands for \textit{Proportional}, \textit{Integral}, and \textit{Derivative}), which is the most common form of feedback control. We design a PID controller to detect data dynamics from the estimates derived by the filtering techniques.   Ultimately, we increase the sampling frequency when data is going through rapid changes and vice versa.

\begin{figure}
\centering
\includegraphics[width=0.9\columnwidth]{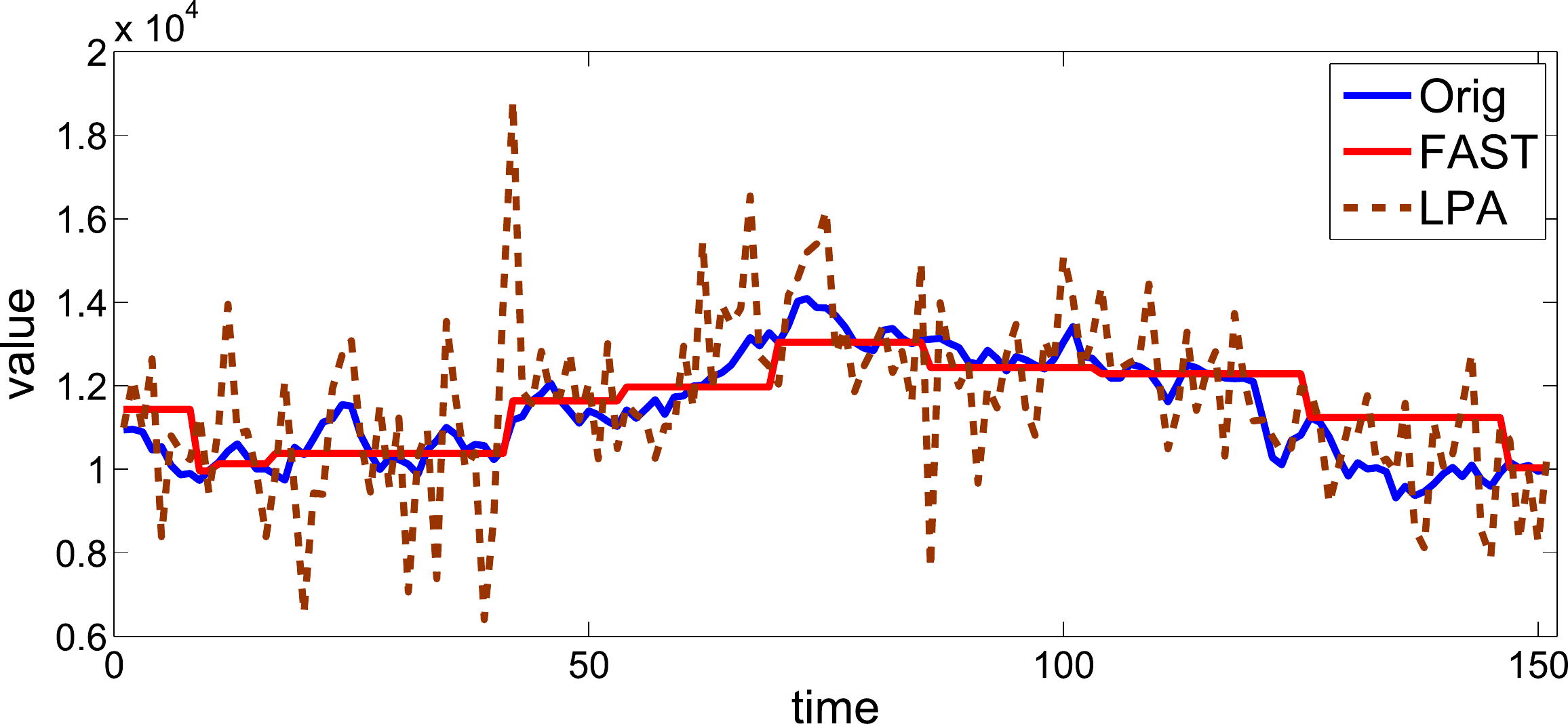}
\vspace{-0.2cm}
\caption{ \small  FAST Released Series with {Linear} Data}
\label{fig:series}
\end{figure}

3) We provide formal analysis on filtering as well as the fixed-rate sampling process to understand their performance.  We empirically evaluate our solution with both real-world and synthetic data sets.  Our experiments show that FAST provides accurate data release and robust performance despite different data dynamics.    Figure~\ref{fig:series} provides an example of the original \textit{linear} times series, the released series by FAST, and that of the baseline LPA algorithm (introduced in Section~3.3).   We observe that FAST released series retains much higher accuracy (i.e. data value, trend) than the LPA released series while providing the same level of privacy guarantee.
With the real-time feature and improved utility, we believe that our solution is applicable to a wider range of monitoring applications.

The rest of the paper is organized as follows: Section II reviews previous works related to differential privacy, time series analysis, and filtering techniques.  Section III provides the problem definition, preliminaries on differential privacy, and outlines of existing solutions. Section IV presents an overview of FAST framework,
as well as the technical details of filtering and sampling.  Section V presents a set of empirical results.   
Section VI concludes the paper and states possible directions for future work.

\section{Related Work}
Here we briefly review the recent, relevant works on differential privacy, time series analysis, and filtering techniques.  

\noindent \textbf{Differential privacy on static data.}  Dwork et al.~\cite{Dwork06} established the guideline to guarantee differential privacy for individual aggregate queries by calibrating the Laplacian noise to the global sensitivity of each query.  Since then, various mechanisms have been proposed to enhance the accuracy of differentially private data release. Blum et al.~\cite{BLR08} proved the possibility of non-interactive data release satisfying differential privacy for queries with
polynomial VC-dimension, such as predicate queries. Dwork et al.~\cite{Dwork09} further proposed more efficient algorithms to release private sanitization of a data set with hardness results obtained.  The work of Hay et al.~\cite{Hay09} improved the accuracy of a tree of counting queries through consistency check, which is done as a post-processing procedure after adding Laplace noise.  This hierarchical structure of queries is referred to as \textit{histograms} by several techniques~\cite{Hay09}\cite{Xiao11TKDE}\cite{Xiao10}\cite{Xu12}\cite{Li10}\cite{Li12}, where each level in the tree is an increasingly fine-grained summary of the data.  
 
A recent study~\cite{Xiao11}, aiming to reduce the relative error, suggests to inject different amount of Laplace noise based on the query result and works well with multidimensional data.  Several other works studied differentially private mechanisms for particular kinds of data, such as search logs~\cite{Korolova09}\cite{Hong12}, sparse data~\cite{Cormode11}, and set-valued data~\cite{Chen11}. When applied to highly self-correlated time-series data, all the above methods, designed to perturb static data,  become problematic because of highly compound Laplace perturbation error.  

\noindent \textbf{Time series analysis and privacy.} Time series data is pervasively encountered in the fields of engineering, science, sociology, and economics. Various techniques~\cite{Brockwell02}, such as ARIMA modeling, exponential smoothing, ARAR, and Holt-Winters methods, have been studied  for time-series forecasting.  Papadimitriou et al.~\cite{Papa07} studied the trade-offs between time-series compressibility property and perturbation.  They proposed two algorithms based on Fast Fourier Transform (FFT) and Discrete Wavelet Transform (DWT) respectively to perturb time-series frequencies.  But the additive noise proposed by them does not guarantee differential privacy, meaning it does not protect sensitive information from adversaries with strong background knowledge.   Rastogi and Nath~\cite{RN10} proposed a Discrete Fourier Transform~(DFT)  based algorithm which implements differential privacy by perturbing the discrete Fourier coefficients.   However, this algorithm cannot produce real-time private released in a streaming environment.  The recent works \cite{ChanSS10}\cite{Dwork10} on continuous data streams defined the \textit{event-level} privacy to protect an event, i.e. one user's presence at a particular time point, rather than the presence of that user.   If one user contributes to the aggregation at time point $t-1$, $t$, and $t+1$, the event-level privacy hides the user's presence at only one of the three time points, resulting the rest two open to attack.

 \noindent \textbf{Filtering.} In our context, "filtering" refers to the derivation of posterior estimates based on a sequence of noisy measurements, in hope of removing background noise from a signal. The Kalman filter, which is based on the use of state-space techniques and recursive algorithms, provides an optimal estimator for linear Gaussian models.   R.E. Kalman published the seminal paper on the Kalman filter~\cite{KALMAN60} in 1960.  Since then, it has become widely applied to areas of signal processing~\cite{Brown97} and assisted navigation systems~\cite{BarShalom02}.  
 It has also gained popularity in other areas of engineering.  One particular application is to wireless sensor networks.  Jain et al.~\cite{Jain04} adopted a dual Kalman filter model on both server and remote sensors to filter out as much data as possible to conserve resources.  Their main concern was to minimize memory usage and communication overhead between sensors and the central server by storing dynamic procedures instead of static data. Increasingly, it has become important to include nonlinearity and non-Gaussianity in order to model the underlying dynamics of a system more accurately.  A very widely used estimator for nonlinear systems is the extended Kalman filter (EKF)~\cite{Sorenson1985} which linearizes about the current mean and error covariance.
 Masreliez~\cite{Masreliez75} proposed solutions to the non-Gaussian filtering problems with linear models in 1975.   His algorithms retain the computationally attractive structure of the Kalman filter but require convolution operations which are hard to implement in all but very simple situations, for instance, when noises can be modeled as a Gaussian mixture.  Gordon et. al~\cite{gordon93} introduced particle filters for solving non-linear non-Gaussian estimation problems in 1993.  Since then, particle methods have become very popular due to the advantage that they do not rely on any local linearisation or any crude functional approximation.  Pitt and Shephard~\cite{Pitt99} introduced auxiliary particle filter and adaptation.  Doucet et al.~\cite{Doucet2000} proposed the optimal importance distribution, approximation, smoothing, and Rao-Blackwellization. A few tutorials, for instance\cite{Arulampalam01atutorial}\cite{Doucet2011},  on particle methods have been published and cover most sequential Monte Carlo algorithms in particle filtering to facilitate implementation.

\section{Preliminaries}

\subsection{Problem Statement}
We formally define an aggregate time series as follows:
\begin{definition}[Aggregate Time Series]
\em A univariate, discrete time series ${\bf X}$ = $\{x_k\}$ is a set of values of an aggregate variable $x$ observed at discrete time $k$,  with $0 \leq k <T$, where $T$ is the length of the series.
\end{definition}

 In our example applications, ${\bf X}$ is an aggregate $count$ series, such as the daily count of patients diagnosed of Influenza, or the hourly count of vehicles passing by a gas station.   This assumption will hold true for the rest of the paper.

\begin{definition}[Utility Metric]
\em We measure the quality of a private, released series ${\bf R}$  = $\{r_k\}$  by average relative error $E$:
\begin{align}
E = \frac{1}{T}\sum_{k}\frac{|r_k - x_k|}{max\{ x_k,\delta\}} 
\end{align}
where $\delta$ is a user-specified constant (also referred to as \textit{sanitary bound} in \cite{Xiao11}) to mitigate the effect of excessively small query results. We set $\delta = 1$ throughout the entire time-series.
\end{definition}

Intuitively, the utility of ${\bf R}$ increases as each $r_k$ approaches $x_k$, the extreme case of which would have $r_k = x_k$ for each $k$.  However, a privacy-preserving algorithm is likely to perturb original data values in order to protect individual privacy.  Thus, our goal is to design a mechanism that guarantees user privacy and yields high utility.

\vspace{-0.1cm}
\subsection{Differential Privacy} 
The privacy guarantee provided by FAST is \textit{differential privacy}~\cite{BLR08}.  Simply put, a mechanism is differentially private if its outcome is not significantly affected by the removal or addition of a single user. 
An adversary thus learns approximately the same information about any individual user, irrespective of his/her
presence or absence in the original database.


\begin{definition}[$\alpha$-Differential Privacy~\cite{BLR08}]
\em A non-interactive privacy 
mechanism $\mathcal{A}$ gives $\alpha$-differential privacy if for any dataset $D_1$
and $D_2$ differing on at most one record, and for any possible anonymized dataset
$\widetilde{D} \in Range(\mathcal{A})$,
\begin{align}
 Pr[\mathcal{A}(D_1)= \widetilde{D}]\leq e^{\alpha}\times Pr[\mathcal{A}(D_2)=
 \widetilde{D}]
\end{align}
\noindent where the probability is taken over the randomness of $\mathcal{A}$. 
\end{definition}

The privacy parameter $\alpha$, also called the \textit{privacy budget}~\cite{McSherry09}, specifies the degree of privacy offered. Intuitively, a lower value of $\alpha$ implies stronger privacy guarantee and a larger perturbation noise, and a higher value of $\alpha$ implies a weaker guarantee while possibly achieving higher accuracy. Two databases $D_1$ and $D_2$ that differ on at most one record are called \textit{neighboring databases}.

\vspace{0.05cm}
\noindent \textbf{Laplace Mechanism.}
Dwork et al.~\cite{Dwork06} show that $\alpha$-differential privacy can be achieved by adding i.i.d. noise $\tilde{N}$ to each query result $q(D)$: 
\begin{align}
\tilde{q}(D) = q(D) + \tilde{N}
\end{align}
 The magnitude of $\tilde{N}$ conforms to a Laplace distribution with probability  $p(x|\lambda)=\frac{1}{2\lambda}e^{-|x|/\lambda}$ and  $\lambda = GS(q) /\alpha$, where $GS(q)$ represents the \textit{global sensitivity}~\cite{Dwork06} of a query $q$.  The global sensitivity is the maximum difference between the query results of $q$ from any two neighboring databases.  In our target applications, each aggregate value is a \textit{count} query and $GS(count) = 1$.  Later on in this paper, we denote the Laplace distribution with $0$ mean and $\lambda$ scale as $Lap(0,\lambda)$.

\vspace{0.05cm}
\noindent \textbf{Composition.}
The composition properties of differential privacy provide privacy guarantees for a
sequence of computations, e.g. a sequence of \textit{count} queries. 

\begin{theorem}[Sequential Composition~\cite{McSherry09}]  \em Let $\mathcal{A}_i$ each provide $\alpha_i$-differential
privacy. A sequence of $\mathcal{A}_i(D)$ over the dataset $D$ provides
($\sum_{i}\alpha_i$)-differential privacy. 
\end{theorem}

\noindent \textbf{User-level privacy vs. Event-level privacy.} The work ~\cite{Dwork10} proposed a differentially private continual counter with the notion of \textit{event}-level privacy, where the neighboring databases differ at $u_i$, a user $u$'s contribution at time stamp $i$.  In our study, we provide a stronger privacy guarantee, \textit{user}-level privacy, where the neighboring databases differ at the user $u$, i.e. $u$'s contribution at all timestamps, thus protecting sensitive information about user $u$ at any time.

\subsection{Existing Solutions}
Here we present the baseline Laplace perturbation algorithm and a recently proposed transformation-based algorithm.  Empirical studies of the two algorithms against our proposed solution are included in Section V.

\begin{algo}
{\bf Input:} Raw series {\bf X}, privacy budget $\alpha$ \\
{\bf Output:} Released series {\bf R} \\ [1.7ex]
\hspace*{2mm}1: {\bf for} each $k \in {0,1,..., T-1}$ {\bf do} \\
\hspace*{2mm}2: \hspace{5mm} $r_k \la$ perturb $x_k$ by $Lap(0,\frac{T}{\alpha})$; \\
\caption{  Laplace Perturbation Algorithm(LPA)} \label{algo:lpa}
\vspace{-0.2 cm}
\end{algo}

\begin{algo}
{\bf Input:} Raw series {\bf X}, privacy budget $\alpha$, parameter $d$\\
{\bf Output:} Released time-series {\bf R} \\ [1.7ex]
\hspace*{2mm}1: compute ${\bf F}^d \la {\bf DFT}^d(X)$ \\
\hspace*{2mm}2: compute $ \widetilde{\bf F}^d \la LPA({\bf F}^d, \alpha)$; \\
\hspace*{2mm}3: compute ${\bf R} \la {\bf IDFT} ({\bf PAD}^T(\widetilde{\bf F}^d))$; \\
\caption{Discrete Fourier Transform(DFT)} \label{algo:dft}
\vspace{-0.2 cm}
\end{algo}

\vspace{0.05in}
\noindent \textbf{Laplace Perturbation Algorithm.} A baseline solution to sharing differentially private time series is to apply the standard Laplace perturbation at each time stamp.  If every query satisfies ${\alpha}/{T}$-differential privacy, by Theorem 1 the sequence of queries guarantees  $\alpha$-differential privacy.    We summarize the baseline algorithm in Algorithm~\ref{algo:lpa} and Line~2 represents the Laplace mechanism to guarantee ${\alpha}/{T}$-differential privacy for each released aggregate.

\vspace{0.05in}
\noindent \textbf{Discrete Fourier Transform.}
Algorithm~\ref{algo:dft} outlines the  Fourier Perturbation Algorithm proposed by Rastogi and Nath~\cite{RN10}.  It begins by computing ${\bf F}^d$, which is composed of the first $d$ Fourier coefficients in the Discrete Fourier Transform~(DFT) of \textbf{X}, with the $j^{th}$ coefficient given as:  $ DFT({\bf X})_j$ $ = \sum_{i=0}^{T-1} e^{\frac{2 \pi \sqrt{-1} }{T} j i} x_i$.  Then it perturbs ${\bf F}^d$ using LPA algorithm with privacy budget $\alpha$, resulting a noisy estimate $ \widetilde{\bf F}^d$.  This perturbation is to guarantee differential privacy.  Denote ${\bf PAD}^T(\widetilde{\bf F}^d)$ the sequence of length $T$ by appending $T-d$ zeros to $\widetilde{\bf F}^d$. The algorithm finally computes the Inverse Discrete Fourier Transform(IDFT) of ${\bf PAD}^T(\widetilde{\bf F}^d)$ to get {\bf R}.  
 The $j^{th}$ element of the inverse is given as: $ IDFT({\bf X})_j = \frac{1}{T} \sum_{i=0}^{T-1} e^{-\frac{2 \pi \sqrt{-1} }{T} j i} x_i$.

The number of coefficients $d$ is a user-specified parameter.  In our empirical study, we set $d=20$ according to the original paper~\cite{RN10}.  As each $IDFT({\bf X})_j$ may be a complex number due to truncation and perturbation, the authors proposed to use $L_1$ distance to measure the quality of the inverse series.  To be consistent,  we adopt the same metric in our empirical study for this algorithm.  We slightly abuse the term and refer to their algorithm as DFT  in the rest of the paper.

\vspace{-0.2cm}
\section{FAST}
We propose FAST, a novel solution to sharing time-series data under differential privacy.  It allows fully automated adaptation to data dynamics and highly accurate time-series prediction/estimation.   
Figure~\ref{fig:fw} illustrates the framework of FAST.  The key steps are outlined below:

\begin{figure}
\centering
\vspace{-0.3cm}
\includegraphics[width=0.95\columnwidth]{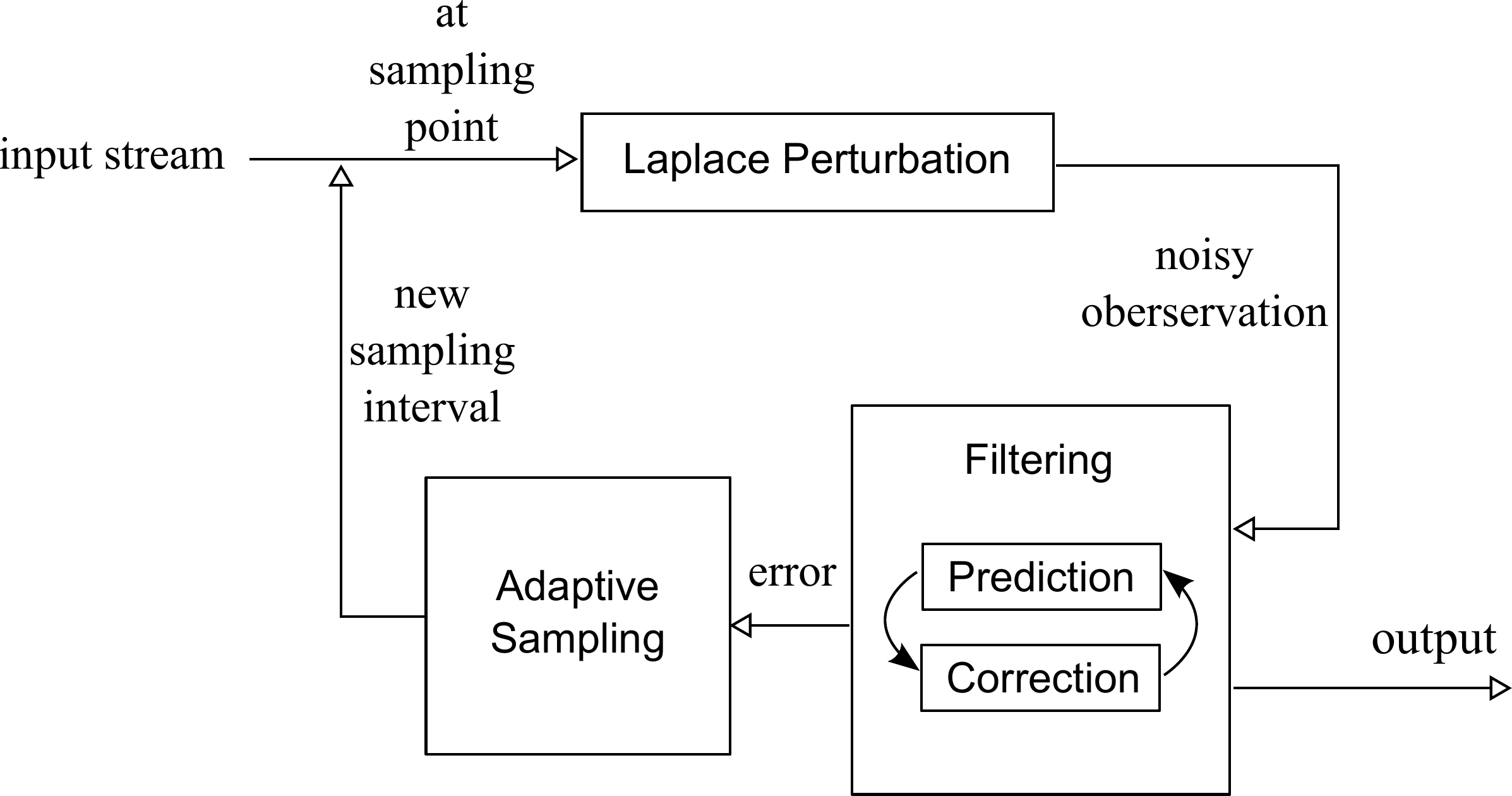}
\caption{\small FAST Framework}
\label{fig:fw}
\end{figure}


\begin{itemize}
\item
For each time stamp $k$, the \textbf{adaptive sampling} component determines whether to sample/query the \textbf{input} time-series or not.  

\item 
If $k$ is a sampling point, the data value at $k$ is perturbed by the {\bf Laplace mechanism} to guarantee differential privacy.   This perturbed value is then received by the \textbf{filtering} component for posterior estimation.  

\item The \textbf{filtering} component produces a $prediction$ of data value based on an internal state model at every time stamp. The prediction, i.e. prior estimate, is released to \textbf{output} at a non-sampling point, while a {\em correction}, i.e. posterior estimate based on both the noisy observation and the prediction, is released at a sampling point.

\item The error between the prior and the posterior estimates is then fed through the \textbf{adaptive sampling} component
 to adjust the sampling rate. Once the user-specified privacy budget $\alpha$ is used up,  the system will stop sampling the input series and will predict at each onward time stamp.
\end{itemize}

\begin{algo}
{\bf Input:} Raw series {\bf X}, privacy budget $\alpha$,  maximum number of samples allowed $M$\\
{\bf Output:} Released series {\bf R} \\ [1.7ex]
\hspace*{2mm}1: {\bf for} each $k$ {\bf do} \\
\hspace*{2mm}2: \hspace{3.5mm} obtain estimate $prior$ from {\bf Prediction}; \\
\hspace*{2mm}3: \hspace{3.5mm} {\bf if} $k$ is a sampling point {\bf \&}  $numSamples < M $ \\
\hspace*{2mm}4: \hspace{6.5mm} $z_k \la$ perturb $x_k$ by $Lap(0,\frac{M}{\alpha})$; \\
\hspace*{2mm}5: \hspace{6.5mm} $numSamples$++;\\
\hspace*{2mm}6: \hspace{6.5mm} obtain estimate $posterior$ from {\bf Correction}; \\
\hspace*{2mm}7: \hspace{6.5mm} $r_k \la posterior$ \\
\hspace*{2mm}8: \hspace{6.5mm} adjust sampling rate by {\bf Adaptive Sampling}; \\
\hspace*{2mm}9: \hspace{3.5mm} {\bf else}  \\
\hspace*{2mm}10: \hspace{5mm}  $r_k \la prior$; \\
\caption{FAST Algorithm} \label{algo:FAST}
\vspace{-0.3 cm}
\end{algo}

Algorithm~\ref{algo:FAST} summarizes our proposed framework.  Given a raw series {\bf X}, overall privacy budget $\alpha$, and maximum number of samples allowed $M$ ($M \le T$), FAST provides real-time estimates of  data values by the \textbf{Prediction} and  \textbf{Correction} procedures and dynamically adjusts the sampling rate with the {\bf Adaptive Sampling} procedure. The details will be discussed in the next two subsections.  Note that FAST evenly distributes the overall privacy budget $\alpha$ to each sample and the exhaustion of $\alpha$ can be detected if $numSamples \ge M$ (Line~3).
\begin{theorem} \em
FAST satisfies $\alpha$-differential privacy. 
\begin{proof}
Given the maximum number of samples $M$  and the overall privacy budget $\alpha$, each sample is $\alpha/M$-differentially private.  According to Theorem 1, Algorithm~\ref{algo:FAST} satisfies $\alpha$-differential privacy.
\end{proof}
\end{theorem}

There are two types of error which we would like to balance in our solution: \textit{perturbation} error by Laplace perturbation mechanism at sampling points and \textit{prediction} error by the filtering prediction procedure at non-sampling points.  
The more we sample, the more \textit{perturbation} error we introduce, while the \textit{prediction} error might be reduced due to more feedback, and vice versa. 
Our goal is to balance the trade-off between the two types of error by adaptively adjusting the sampling rate.  

\vspace{-0.15cm}
\subsection{Filtering}
The filtering component in FAST generates estimates of monitored aggregates in order to improve the quality of released data per time stamp.   We first establish below a state-space model for the time series data of the example applications in Section I.   Later in this section, we propose and present the details of two filtering algorithms for estimation.     

\vspace{0.05in}
\noindent \textbf{Process Model.} Let  $x_k$ denote the internal state, i.e. true value, of a process at time stamp $k$. The states at consecutive time stamps can be modeled by the following equations:
\begin{align}
x_k = x_{k-1} + \omega \\
\omega \sim N(0, Q)
\end{align}
This constant process model indicates that adjacent data values from the original time-series should be consistent except for a white Gaussian noise $\omega$, called the process noise, with variance $Q$ .

\vspace{0.05in}
\noindent \textbf{Measurement Model.} The noisy observation, which is provided by the Laplace mechanism,  can be represented as follows:
\begin{align}
z_k = x_k + \nu \\
\nu \sim Lap(0, 1/\alpha_0)
\end{align}
where $\nu$ is called the measurement noise.  Clearly, the noisy observation $z_k$ is the true state plus the perturbation noise.   $\alpha_0$ denotes the differential privacy budget for each sample, e.g. $\alpha_0=\alpha/T$ if sampling at every time stamp; $\alpha_0 = \alpha/M$ if no more than $M$ samples are allowed.

\vspace{0.05in}
\noindent \textbf{Posterior Estimate.} 
With the noisy observation $z_k$ at time stamp $k$, we propose to release the \textit{aposteriori} estimate of the true state $x_k$, in order to improve the utility of the released value.  The posterior estimate, denoted by $\hat{x}_k$, can be given by the following conditional expectation:
\begin{align}
\hat{x}_k = E(x_k|\mathbb{Z}^k)
\end{align}
where $\mathbb{Z}^k = \{z_0, z_1,...,z_k\}$ denotes the set of observations obtained so far. Therefore, we can derive $\hat{x}_k$ only if the \textit{aposteriori} probability density function $f(x_k|\mathbb{Z}^{k})$ can be determined.  According to Bayes' theorem, we obtain the following relation between two consecutive time stamps:
\begin{align}
f(x_k|\mathbb{Z}^{k}) = \frac{f(x_k|\mathbb{Z}^{k-1})f(z_k|x_k)}{f(z_k|\mathbb{Z}^{k-1})},
\end{align}
where the prior and the normalizing constant are given by:
\begin{align}
f(x_k|\mathbb{Z}^{k-1}) = \int f(x_{k-1}|\mathbb{Z}^{k-1})f(x_k|x_{k-1})dx_{k-1},\\
f(z_k|\mathbb{Z}^{k-1}) = \int f(x_k|\mathbb{Z}^{k-1}) f(z_k|x_k) dx_k.
\end{align}
In general, Equations~(9-11) are difficult to carry out when $f(z_k|x_k)$, i.e. $f(\nu = z_k-x_k)$, is non-Gaussian.  Therefore, the posterior density cannot be analytically determined without the Gaussian assumption about the measurement noise.  

We propose two solutions to the posterior estimation challenge discussed above.  One is to approximate the Laplace noise with a Gaussian noise; the other is to simulate the posterior density function with Monte Carlo methods.  The details are presented below, respectively.  

\subsubsection{Gaussian Approx. of Measurement Noise}

In our previous work~\cite{Fan12CIKM}, we propose to approximate the Laplace noise $\nu$ with a white Gaussian error
\begin{equation}
\nu \sim N(0, R)
\end{equation} 
 and therefore the estimation of $\hat{x}_k$ in Equation~(8) can be solved with the classic Kalman filter~\cite{KALMAN60}.

\vspace{0.05in}
\noindent \textbf{The Kalman Filter}. At time stamp $k$, the prior state estimate $\hat{x}_k ^ -$ is made according to the process model in Equation~(4) and is related to the posterior estimate of the previous step: 
\begin{align}
\hat{x}_k^- = \hat{x}_{k-1}.
\end{align}
The posterior estimate $\hat{x}_k$ can be given as a linear combination of the prior $\hat{x}_k^-$ and the observation $z_k$:
\begin{align}
\hat{x}_k = \hat{x}_k^- + K_k( z_k - \hat{x}_k^-).
\end{align}
where $K_k$, the \textit{Kalman Gain}, is adjusted at every time stamp to minimize the posterior error variance.  Below we briefly show how $K_k$  is derived for each time stamp $k$.

Let $P_k^-$ and $P_k$ denote the \textit{apriori} and \textit{aposteriori} error variance, respectively.   They are defined as
\begin{align}
P_k^- = E [ (x_k - \hat{x}_k^- )(x_k - \hat{x}_k^- )^T] \\
P_k = E [ (x_k - \hat{x}_k )(x_k - \hat{x}_k )^T]
\end{align}

By the Gaussian assumption regarding $\omega$ and $\nu$ and given the prior error variance $P_k^-$ at time stamp $k$, we can substitute Equation~(14, 15)  into Equation~(16) and apply the gradient descendant method to minimize $P_k$.  Therefore, we obtain an optimal value for $K_k$ as
\begin{align}
K_k=P_k^- ( P_k^-  + R)^{-1}
\end{align}
and thus the optimal $P_k$ is
\begin{align}
P_k= (1 - K_k)P_k^- .
\end{align}
Similarly, given $P_k$, we can easily project the prior error variance at $k+1$ according to Equation~(4):
\begin{align}
P_{k+1}^- = P_k + Q .
\end{align}

The classic Kalman filter recursively performs two operations:  \textit{Prediction} and \textit{Correction}, which correspond to prior and posterior estimation respectively.  Algorithm~\ref{algo:KFP} and \ref{algo:KFC} provide details of the two estimation steps used in FAST framework.

\begin{algo}
{\bf Input:} Previous release $r_{k-1}$  \\
{\bf Output:} Prior estimate $\hat{x}_k^-$ \\ [1.7ex]
\hspace*{2mm}1: $\hat{x}_k^- \la r_{k-1}$; \\
\hspace*{2mm}2: $P_k^- \la P_{k-1} + Q; $ \\
\caption{KFPredict($k$)} \label{algo:KFP}
\end{algo}

\begin{algo}
{\bf Input:} Prior $\hat{x}_k^-$ , noisy measurement $z_k$\\
{\bf Output:} Posterior estimate $\hat{x}_k$ \\ [1.7ex]
\hspace*{2mm}1: $K_k \la P_k^- (P_k^- + R)^{-1}$; \\
\hspace*{2mm}2: $\hat{x}_k \la \hat{x}_k^- + K_k(z_k-\hat{x}_k^-); $ \\
\hspace*{2mm}3: $P_k \la (1-K_k)P_k^- ;$\\
\caption{KFCorrect($k$)} \label{algo:KFC}
\end{algo}

\vspace{0.05in}
\noindent \textbf{Accuracy}.  Here we study the performance of the Kalman filter based algorithm without sampling.  Therefore, a noisy observation is obtained and the \textit{Prediction} and \textit{Correction} pair is performed at every time stamp. The Kalman filter is optimal only when the Gaussian assumption regarding the measurement noise holds, i.e. Equation~(12).  We analyze the posterior error variance $var({x}_k - \hat{x}_k)$ where $\hat{x}_k$ is given by Algorithm~\ref{algo:KFC} but $z_k$ is obtained from the Laplace Mechanism, i.e. Equation~(6,7), which violates the Gaussian assumption. 
The goal is to find the optimal approximate Gaussian noise, i.e. the optimal value of  $R$,  in order to achieve minimum variance posterior estimate.  Due to the recursive nature of filtering, it's difficult to obtain a closed form for the optimal $R$ value.  We conclude our main finding in the theorem below and refer readers to Appendix A for the detailed least square analysis.  

\begin{theorem}[Optimal Approximation]  \em Given an original data series of length $T$ and the overall privacy budget $\alpha$,  using an approximate Gaussian noise $R$ leads to the following posterior error:
\begin{align}
var({x}_k - \hat{x}_k) =  \nonumber \\
&  \hspace{-2.5cm} \frac{ R^2 [ var({x}_{k-1} - \hat{x}_{k-1})  + Q]}{(P_k^- + R)^2} +\frac{2{P_k^- }^2 T^2}{(P_k^- + R)^2 \alpha^2}
\end{align}
and optimal posterior estimation requires $R \propto \frac{T^2}{\alpha^2}$.
\begin{proof}
See Appendix A.
\end{proof}
\end{theorem}

Theorem~3 provides guidance for choosing the Gaussian measurement noise, i.e. the value of $R$, to approximate the perturbation noise introduced by differential privacy mechanism. In this case,  the perturbation noise conforms to $Lap(0, 1/\frac{\alpha}{T})$. The result confirmed that the optimal $R$ is proportional to the variance of the Laplace perturbation noise, given the length of the series and the overall privacy budget. 

\vspace{0.05in}
\noindent \textbf{Estimation with Sampling}. Note that the posterior estimate $\hat{x}_k$ cannot be determined when noisy observation $z_k$ is absent.  When combined with sampling in our overall solution, we propose to estimate as follows: at sampling points, i.e. when noisy observations are available, both \textit{Prediction} and \textit{Correction} will be performed and the posterior estimates will be released; at non-sampling points, i.e. when noisy observations are absent, only the \textit{Prediction} step will be performed and the prior estimates will be released.

One advantage of the Kalman filter is that  it maintains and updates the best estimate of the internal state by properly weighing and combining all available data (prior and noisy observation). Another advantage is the computation efficiency which is crucial to real-time applications,  as only $O(1)$ computations are required for each time stamp from Algorithm~\ref{algo:KFP} and \ref{algo:KFC}. 

\subsubsection{Monte Carlo Estimation of Posterior Density}
Without approximating the Laplace noise with a Gaussian error,  Monte Carlo methods can be used to represent the posterior density function $f(x_k|\mathbb{Z}^{k})$ by simulation.  In this section, we will show the solution to posterior estimation based on the Sampling-Importance-Resampling(SIR) particle filter, which is also known as bootstrap filtering~\cite{gordon93} and condensation algorithm~\cite{maccormick99}. 



\vspace{0.05in}
\noindent \textbf{SIR Particle Filtering}. With a collection of $N$ weighted samples or particles, $\{x_k^i, \pi_k^i\}_{i=1}^N$, where $\pi_k^i$ is the weight of particle $x_k^i$, the posterior density at timestamp $k$ can be represented as follows:
\begin{equation}
f(x_k|\mathbb{Z}^{k}) = \sum_{i=i}^{N} \pi_k^i \delta(x_k - x_k^i)
\end{equation}
where $\delta(\cdot)$ is Dirac delta measure.  

The weights  $\{\pi_k^i\}_{i=1}^N$ are chosen according to the importance sampling method, 
where particles  $\{x_k^i\}_{i=1}^N$ can be easily generated from a proposal $q(\cdot)$ called an \textit{importance density}.   
The details of the importance sampling method are omitted here and can be found in \cite{Doucet2001}.  By assuming that the importance density $q(\cdot)$ depends only on the previous state and current measurement, the following weight relationship between two successive timestamps can be derived:
\begin{equation}
\pi_k^i \propto \pi_{k-1}^i \frac{ f(z_k|x_k^i) f(x_k^i|x_{k-1}^i)}{q(x_k^i|x_{k-1}^i,z_k)} .
\end{equation} 
According to Arulampalam et al~\cite{Arulampalam01atutorial}, it is offen convenient to choose the importance density $q(\cdot)$ to be the prior density $f(x_k|x_{k-1}^i)$. Substituting it into Equation~(22)  then yields
\begin{equation}
\pi_k^i \propto \pi_{k-1}^i f(z_k|x_k^i) 
\end{equation}
where
\begin{equation}
 f(z_k|x_k^i) = f(\nu = z_k - x_k^i)
\end{equation}
and $\nu$ is the Laplace noise as defined in Equation~(7).

The SIR particle filter explicitly employs a resampling step at every time stamp in order to circumvent degeneracy phenomenon, where after a few iterations, all but one particle will have negligible weight.  In our solution, we adopt systematic resampling as recommended in~\cite{Arulampalam01atutorial}. 
Since $\pi_{k-1}^i = 1/N$ for every $i$ after resampling, weights at time $k$ can be simplified as follows:
\begin{equation}
\pi_k^i \propto  f(z_k|x_k^i).
\end{equation}
We will use the above result for correction in the overall algorithm.

\begin{algo}
{\bf Input:}  Particles at time $k-1$ $\{x_{k-1}^i, \pi_{k-1}^i\}_{i=1}^N$\\
{\bf Output:} Prior estimate $\hat{x}_k^-$ \\ [1.7ex]
\hspace*{2mm}1: {\bf for} each $i \in {1,...,N} $ {\bf do} \\
\hspace*{2mm}2: \hspace{3mm} draw $x_k^i \sim f(x_k|x_{k-1}^i)$ \\
\hspace*{2mm}3: $\hat{x}_k^- \la \frac{1}{N} \sum_{i=1}^{N} x_k^i$ \\
\caption{ PFPredict(k)} \label{algo:PFPredict}
\vspace{-0.15 cm}
\end{algo}

\begin{algo}
{\bf Input:}  Particles $\{x_{k}^i\}_{i=1}^N$, noisy measurement $z_k$\\
{\bf Output:} Posterior estimate $\hat{x}_k$\\ [1.7ex]
\hspace*{2mm}1: {\bf for} each $i \in {1,...,N} $ {\bf do} \\
\hspace*{2mm}2: \hspace{3mm} assign particle weight $\pi_k^i$ according to $(25)$ \\
\hspace*{2mm}3:  normalize $\{\pi_{k}^i\}_{i=1}^N$\\
\hspace*{2mm}4: $\hat{x}_k \la \sum_{i=1}^{N} \pi_k^i x_k^i$ \\
\hspace*{2mm}5: resample from $\{x_k^i, \pi_k^i\}_{i=1}^N$\\
\caption{ PFCorrect(k)} \label{algo:PFCorrect}
\vspace{-0.15 cm}
\end{algo}

\vspace{0.05in}
\noindent \textbf{Prediction and Correction}. Algorithm~\ref{algo:PFPredict} and \ref{algo:PFCorrect} provide details of  the particle filtering based estimation algorithm used in FAST framework. 

For each particle $i$, the \textit{Prediction} step  (Line~1-2 in Algorithm~\ref{algo:PFPredict}) projects its value for the next time stamp according to the process model $f(x_k|x_{k-1}^i)$.  Note that $f(x_k|x_{k-1}^i)$  represents the distribution $N(x_{k-1}^i,Q)$, according to Equation~(4).  
Once the particles are drawn, the prior estimate can be given with the uncorrected weights (Line~3 in Algorithm~\ref{algo:PFPredict}):
\begin{equation} 
\hat{x}_k^- = \sum_{i=1}^{N} \pi_{k-1}^i x_k^i = \frac{1}{N} \sum_{i=1}^{N} x_k^i.
\end{equation} 

 The \textit{Correction} step adjusts particle weights according to the noisy observation $z_k$.
After weight adjustment and normalization (Line~1-3 in Algorithm~\ref{algo:PFCorrect}), the posterior estimate can be derived as follows (Line~4 in Algorithm~\ref{algo:PFCorrect}):
\begin{equation}
\hat{x}_k = \sum_{i=1}^{N} \pi_k^i x_k^i .
\end{equation}
As implied by the SIR particle filtering method, resampling is applied at the end of the \textit{Correction} step (Line~5 in Algorithm~\ref{algo:PFCorrect}).

The initialization of particles $\{x_0^i, \pi_0^i\}_{i=1}^N$ is non-trivial, since the distribution of the initial state is unlikely to be available in many real-time applications.  Therefore, we skip the estimation steps, i.e. Equation~(26-27), at time $0$ and release the noisy measurement $z_0$.  Particles $\{x_0^i\}_{i=1}^N$ are then uniformly drawn from the vicinity of $z_0$ and $\{\pi_0^i\}_{i=1}^N$ are initialized as $1/N$.

\vspace{0.05in}
\noindent \textbf{Accuracy}. We refer the readers to \cite{Doucet2011}  for the accuracy and convergence results of the SIR algorithm.  Intuitively, a larger number of particles implies a more accurate distribution estimation.  On the other hand, a larger number of particles requires more computation time, which is crucial to real-time monitoring applications. As a matter of fact, the complexity of Algorithm~\ref{algo:PFPredict} and~\ref{algo:PFCorrect} is $O(N)$ per time stamp.   We will examine the trade-off between accuracy and run time of Algorithm~\ref{algo:PFPredict} and~\ref{algo:PFCorrect} in the experiment section.

\vspace{0.05in}
\noindent \textbf{Estimation with Sampling}.  Combined with sampling in the overall solution,  the particle filtering based algorithm adopts the same strategy as the Kalman filter: it releases  posterior estimates at sampling points and  prior estimates at non-sampling points.    The utility of the particle filtering based algorithm will be evaluated against other methods in Section V.

\vspace{-0.15cm}
\subsection{Sampling}
Since each noisy observation from Laplace mechanism comes with a cost (privacy budget spent), we are motivated to sample data values through the differential privacy interface only when needed in our overall solution.   Below we propose two sampling strategies: one is to sample the series with a fixed interval, while the other is to dynamically adapt the sampling rate based on feedback control.

\vspace{+0.1cm}
\noindent \textbf{Fixed Rate Sampling.}
Given a pre-defined interval $I$, the fixed-rate algorithm samples the time series periodically and releases the posterior estimate per $I$ timestamps.  As for the time points between two adjacent samples, a predicted value/prior estimate is released.   Privacy budget $\alpha/(\frac{T}{I})$ will be spent on each sample to guarantee $\alpha$-differential privacy for the entire series according to Theorem 1.

\begin{algo}
{\bf Input:}  Current time stamp $k$, fixed interval $I$\\
{\bf Output:} Sampling or not \\ [1.7ex]
\hspace*{2mm}1: {\bf if} $k \% I == 0$ \\
\hspace*{2mm}2: \hspace{3mm} $k$ is a sampling point \\
\hspace*{2mm}3: {\bf else} \\
\hspace*{2mm}4: \hspace{3mm} $k$ is a non-sampling point \\
\caption{Fixed Rate Sampling} \label{algo:fixed}
\vspace{-0.15 cm}
\end{algo}

Algorithm~\ref{algo:fixed} summarizes the fixed rate sampling algorithm which can be used in FAST framework.  The challenge of fixed-rate sampling is to determine the optimal interval $I$.  When increasing the sampling rate, i.e. when $I$ is low, an extreme case of which is to issue a query at each time step as in the baseline solution, the perturbation error introduced at each time stamp is increased. On the other hand, when we decrease the sampling rate, i.e. when $I$ is high, the perturbation at each sampling point will drop, but the published series will not reflect up-to-date data values, resulting large prediction error.  We analyze the posterior error of fixed-rate sampling and find that it is very challenging to quantify and minimize the sum of error \textit{a priori}.  Detailed discussion is in Appendix B.  Therefore, the fixed-rate sampling method may not be optimal in our problem setting.

\begin{figure}
\centering
\includegraphics[width=0.8\columnwidth]{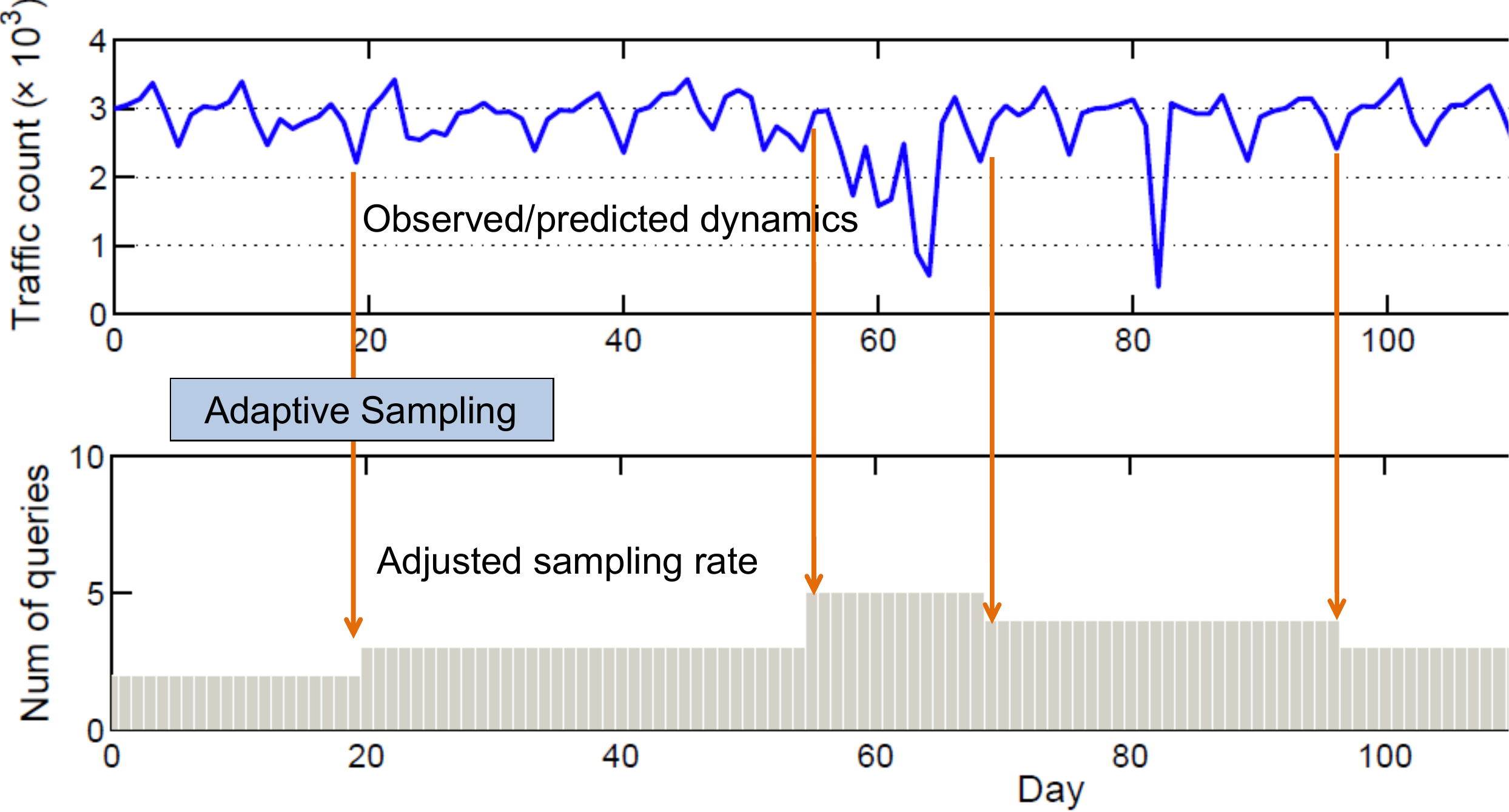}
\vspace{-0.35cm}
\caption{ \small Adaptive Sampling with {Traffic} Data}
\label{fig:adaptive_sampling}
\end{figure}

\vspace{+0.1cm}
\noindent \textbf{Adaptive Sampling.}
With no \textit{a priori} knowledge of the time series, it is desirable to detect data dynamics and to adjust the sampling rate on-the-fly. 
Figure~\ref{fig:adaptive_sampling} illustrates the idea of adaptive sampling. We plot the original \textit{traffic} series  as well as the number of queries (samples) issued by the adaptive sampling mechanism during each corresponding time unit.
 As is shown, the adaptive sampling mechanism increases sampling rate between day 20 and day 100, when the traffic count exhibits significant fluctuations, and decreases sampling rate beyond day 100, when there's little variation among data values.

In FAST framework, we propose adaptive sampling with feedback control.  It is based on the error between the posterior and the prior estimates from the filtering module, which is defined below.

\begin{definition}[Feedback Error]
\em At time step $k_n$ ($0\le k_n<T$), where the subscript $n$ indicates the $n$-th sampling point ($0 \le n<M$), we define the feedback error $E_{k_n}$ as follows:
\vspace{-0.1cm}
\begin{align}
E_{k_n} = |\hat{x}_{k_n} - \hat{x}_{k_n}^-|/\max\{\hat{x}_{k_n}, \delta\}.
\vspace{-0.1cm}
\end{align}
Note that no error is defined at a non-sampling point. 
\end{definition}

The feedback error measures how well the internal state model describes the current data dynamics, assuming $\hat{x}_{k_n}$ is close to the true state.  Since $\hat{x}_{k_n}^-$ is given by a constant state model, we may infer that data is going through rapid changes when the error $E_{k_n}$ increases.  In response, the controller in our system will detect the error and increase the sampling rate accordingly. 

FAST adopts a PID controller, the most common form of feedback control~\cite{king2011},  to measure the performance of  sampling over time.  We re-define the three PID components, \textit{Proportional, Integral}, and \textit{Derivative}, with the feedback error defined in Equation~(28).
\begin{itemize}

\item {\bf Proportional} error is to keep the controller output ($\Delta$) in proportion to the current error $E_{k_n}$ with $k_n$ being the current time step and subscript $n$ being the sampling point index. It is given by $ \Delta_p = C_p  E_{k_n}$ where $C_p$ denotes the \textit{proportional} gain which amplifies the current error.  

\item {\bf Integral} error is to eliminate offset by making the change rate of control output proportional to the error.   With similar terms, we define the integral control as $\Delta_i = \frac{C_i}{T_i}\sum_{j=n-T_i+1}^n E_{k_j}$
where $C_i$ denotes \textit{integral} gain amplifying the integral error and $T_i$ represents the integral time window indicating how many recent errors are taken.

\item {\bf Derivative} error attempts to prevent large errors in the future by changing the output in proportion to the change rate of error. It is defined as $\Delta_d = {C_d} \frac{E_{k_n} -  E_{k_{n-1}}}{k_n - k_{n-1}}$
where $C_d$ is \textit{derivative} gain amplifying the derivative error.

\end{itemize}
The full PID algorithm is thus
\begin{equation}
\Delta=  C_p  E_{k_n}+  \frac{C_i}{T_i}\sum_{j=n-T_i+1}^n E_{k_j}+  {C_d} \frac{E_{k_n} -  E_{k_{n-1}}}{k_n - k_{n-1}}.
\end{equation}
Control gains $C_p$, $C_i$, and $C_d$ denote how much each of the \textit{proportional}, \textit{integral}, and \textit{derivative} counts for the final calibrated PID error.  In FAST, they are constrained by:
\vspace{-0.15cm}
\begin{align}
C_p, C_i, C_d \ge 0 \\
C_p + C_i + C_d = 1.
\end{align}
Note that setting $C_i > 0$ requires $T_i$ previous samples in order to evaluate the integral error, which can be implemented as a straight-forward initialization prior to adaptive adjustment of the sampling rate.

\begin{algo}
{\bf Input:}  Current time stamp $k$, next sampling point $ns$ \\
{\bf Output:} Sampling or not \\ [1.7ex]
\hspace*{2mm}1: {\bf if} $k == ns$ \\
\hspace*{2mm}2: \hspace{3mm} $k$ is a sampling point \\
\hspace*{2mm}3: \hspace{3mm} afterwards, obtain feedback from {\bf filtering} \\
\hspace*{2mm}4: \hspace{3mm} update $\Delta$ according to (29) \\
\hspace*{2mm}5: \hspace{3mm} calculate $I'$ according to (32) \\
\hspace*{2mm}6: \hspace{3mm} $ns \la ns + I'$, $I \la I'$ \\
\hspace*{2mm}7: {\bf else} \\
\hspace*{2mm}8: \hspace{3mm} $k$ is a non-sampling point \\
\caption{Adaptive Sampling} \label{algo:adaptive}
\vspace{-0.15 cm}
\end{algo}

Given the PID error $\Delta$, a new sampling interval $I'$ can be determined:
\vspace{-0.1cm}
\begin{equation}
I' = max\{1, I + \theta(1- e^{\frac{\Delta - \xi}{\xi}})\}
\end{equation} 
where $\theta$ and $\xi$ are user-specified parameters.  By default, the smallest sampling interval is set to $1$.   Parameter $\theta$ determines the magnitude of change and $\xi$ is the set point for the sampling process.  
We assume $\xi$ is $10\%$ in our empirical studies, i.e. the maximum tolerance for PID error is $10\%$.  It can be determined by the users according to specific applications.     

Algorithm~\ref{algo:adaptive} summarizes the adaptive sampling algorithm used in FAST framework.  It maintains and updates the variable $ns$ indicating the next sampling point.  If the current time stamp is determined to be a sampling point, a feedback error can be obtained from the filtering component (Line~3) after correction.  The current PID error $\Delta$ can then be evaluated (Line~4) as well as a new sampling interval $I'$ (Line~5).  A future sampling point, i.e. updated $ns$, is derived by applying the new sampling interval $I'$ (Line~6).  When $k$ is a non-sampling point, the algorithm receives no feedback since only the prediction step is run in the filtering component.  We will study the parameters as well as the performance of both sampling algorithms in the next section. 

\vspace{-0.1cm}
\section{Experiment}

We have implemented FAST in Java with JSC (Java Statistical Classes\footnote{http://www.jsc.nildram.co.uk}) for simulating the statistical distributions.   
Our study has been conducted with synthetic as well as real-world  data sets.  The synthetic data sets are 1000 time stamps long and generated with $Q=10^5$ to incorporate data value fluctuations.   

\begin{itemize}
\item \textbf{Linear} is a synthetic series generated by our process model in Equation~(4).
\item \textbf{Logistic} is a synthetic series generated by the logistic model $x_k = A(1+e^{-k})^{-1}$ with $A=5000$.  
\item \textbf{Sinusoidal} is a synthetic series generated by a sinusoid $x_k = A\sin(bk+c)$ with $A=5000, b=\pi/6, c=\pi/2$.
\end{itemize}

The real-world data sets are of variable length.
\begin{itemize}
\item \textbf{Flu} is the weekly surveillance data of Influenza-like illness provided by the Influenza Division of the Centers for  Disease Control and Prevention\footnote{http://www.cdc.gov/flu/}.   We collected the weekly outpatient count of the age group [5-24] from 2006 to 2010.  This time-series consists of 209 data points.
\item \textbf{Traffic} is a daily traffic count data set for Seattle-area highway traffic monitoring and control provided by the Intelligent Transportation Systems Research Program at University of Washington\footnote{http://www.its.washington.edu/}.   We chose the traffic count at location I-5 143.62 southbound from April 2003 till October 2004.  This time-series consists of 540 data points.
\item \textbf{Unemploy} is the monthly unemployment level of black or African American women of age group [16-19] from ST. Louis Federal Reserve Bank\footnote{http://research.stlouisfed.org/}.  This data set contains observations from January 1972 to October 2011 with 478 data points.
\end{itemize}

\begin{figure}
\centering
 \subfigure[\textit{flu} data]{ 
    \label{fig:kalman_R:a} 
    \includegraphics[width=0.41\columnwidth]{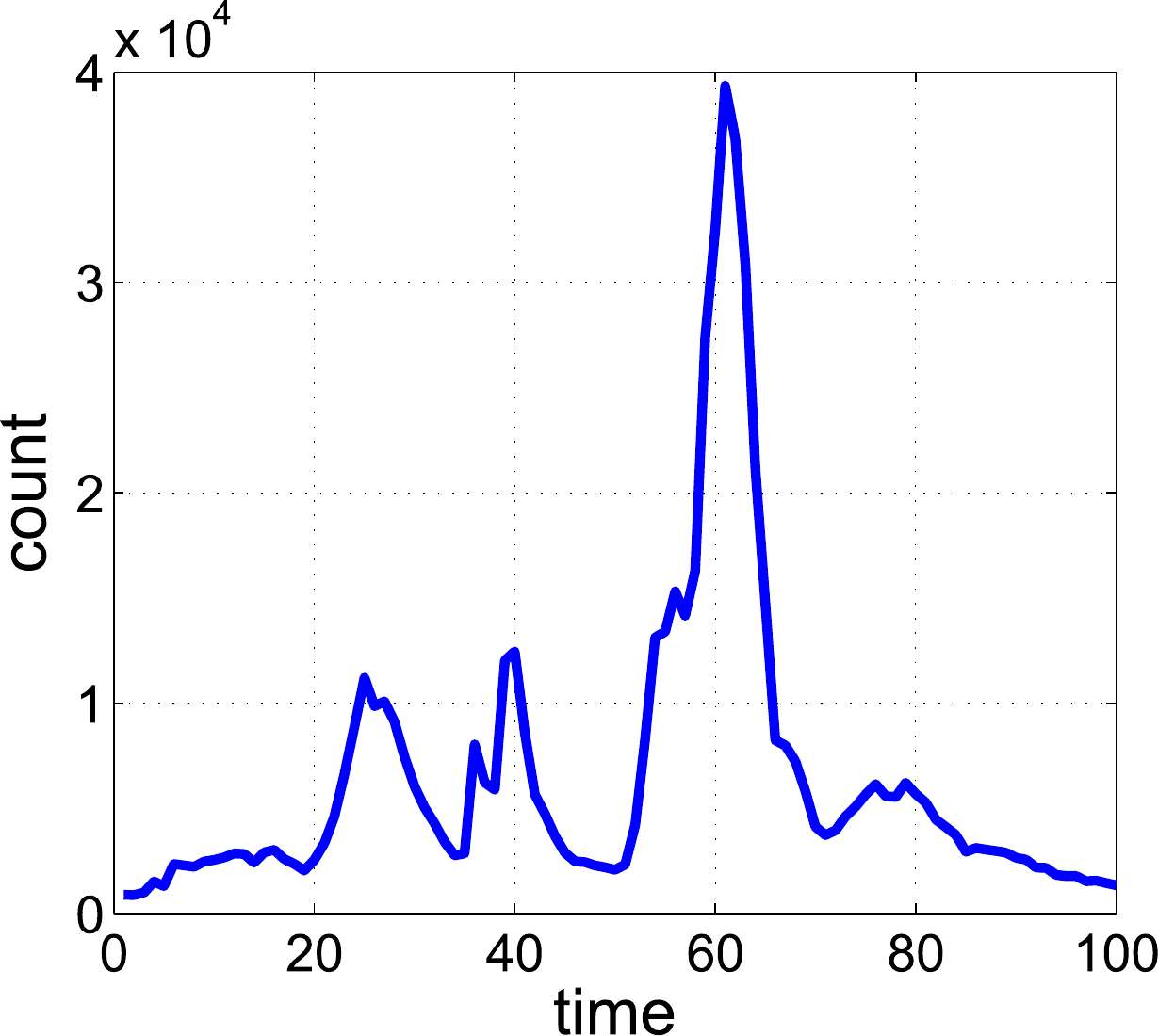}} 
  \subfigure[\textit{traffic} data]{ 
    \label{fig:kalman_R:b} 
    \includegraphics[width=0.45\columnwidth]{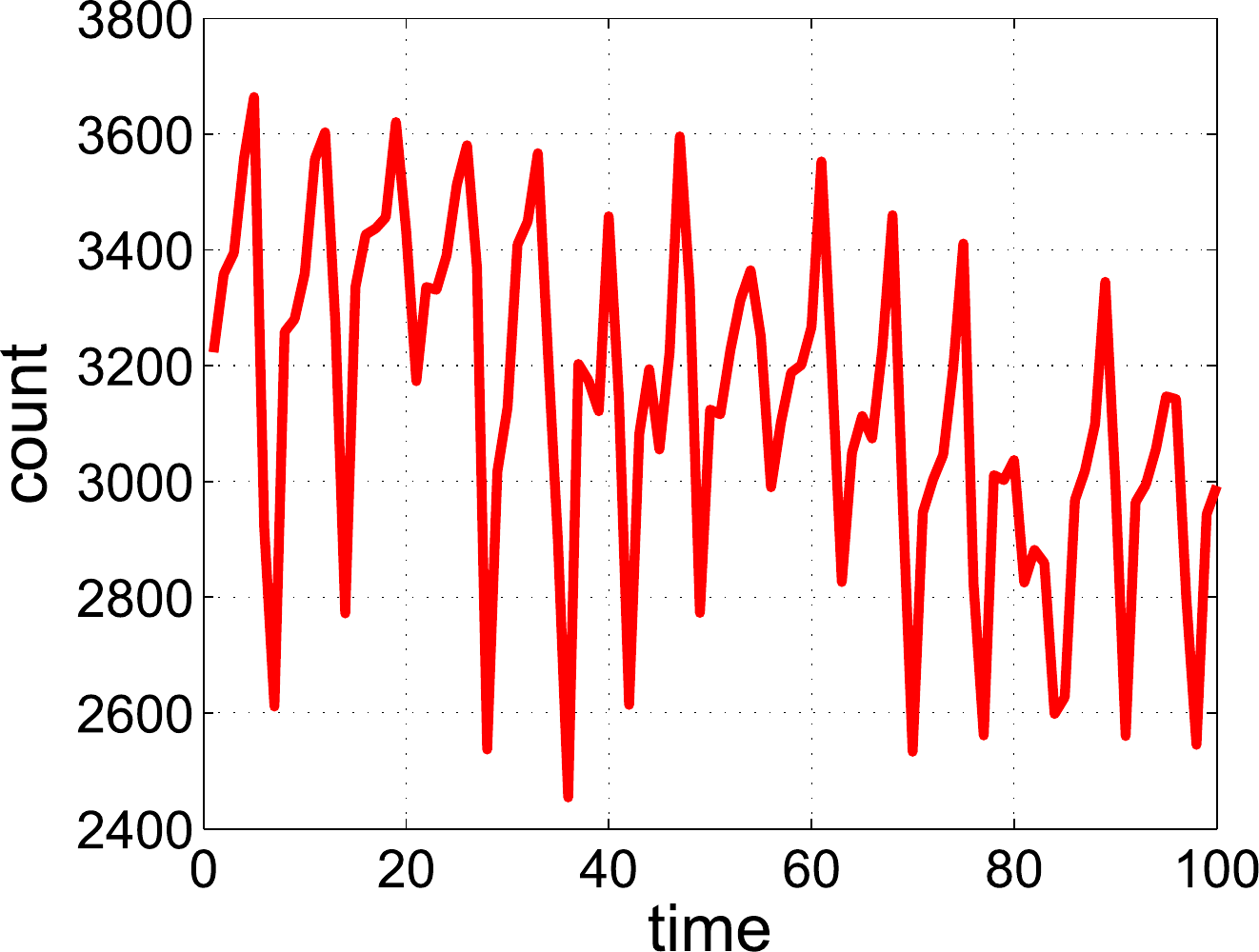}} 
  \caption{\small Data Snapshots} 
\label{fig:data}
\end{figure}

Figure~\ref{fig:data} illustrates the different dynamics of the data sets.  For instance, the \textit{flu} data set has a relatively smooth curve and reflects significant changes in data value, while the \textit{traffic} data has a less smooth curve but fluctuates around a rather stable average value.

To show the impact of FAST parameters, we will only present empirical results with the \textit{Linear} data set for brevity.  The default parameter setting, unless otherwise noted, is summarized in Table~\ref{tab:param}. 

\begin{table}[t]
\centering
\caption{FAST Parameters}
\begin{tabular}{|c|l|c|} \hline
{\bf Symbol}&{\bf Description}&{\bf Default Value}\\ \hline
$\alpha$ & Total Privacy Budget & 1\\ \hline
$Q$ & Process Noise & $10^5$\\ \hline
$R$ & Gaussian Measurement Noise & $10^6$\\ \hline
$N$ & Number of Particles & $10^3$\\ \hline
$(C_p, C_i,C_d)$ & Control Gains & $(0.9, 0.1, 0)$\\ \hline
$T_i$ & Integral Time Window & $5$\\ \hline
$(\theta,\xi)$ & Interval Adjustment Params & $(10, 0.1)$\\ \hline
\end{tabular}\label{tab:param}
\end{table}

\subsection{Effects of Filtering}
Here we study the impact of parameters on filtering performance alone. Therefore,  no sampling is applied in the experiments of this section, hence a posterior estimate can be derived for each time stamp. 

\begin{figure}
\centering
 \subfigure[$R$ vs. $T$]{ 
    \label{fig:kalman_R:a} 
    \includegraphics[width=0.45\columnwidth]{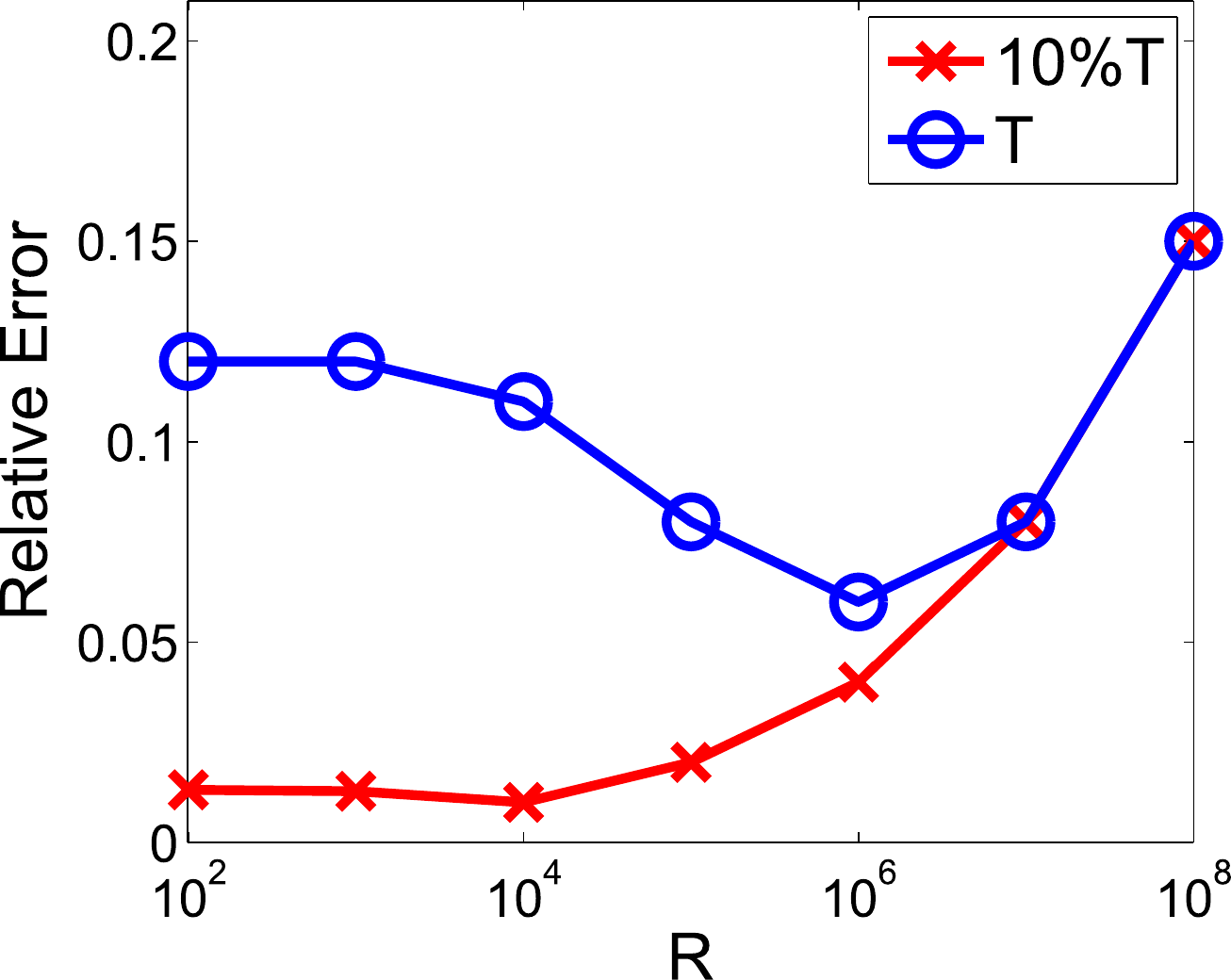}} 
  \hspace{0.02in} 
  \subfigure[$R$ vs. $\alpha$]{ 
    \label{fig:kalman_R:b} 
    \includegraphics[width=0.45\columnwidth]{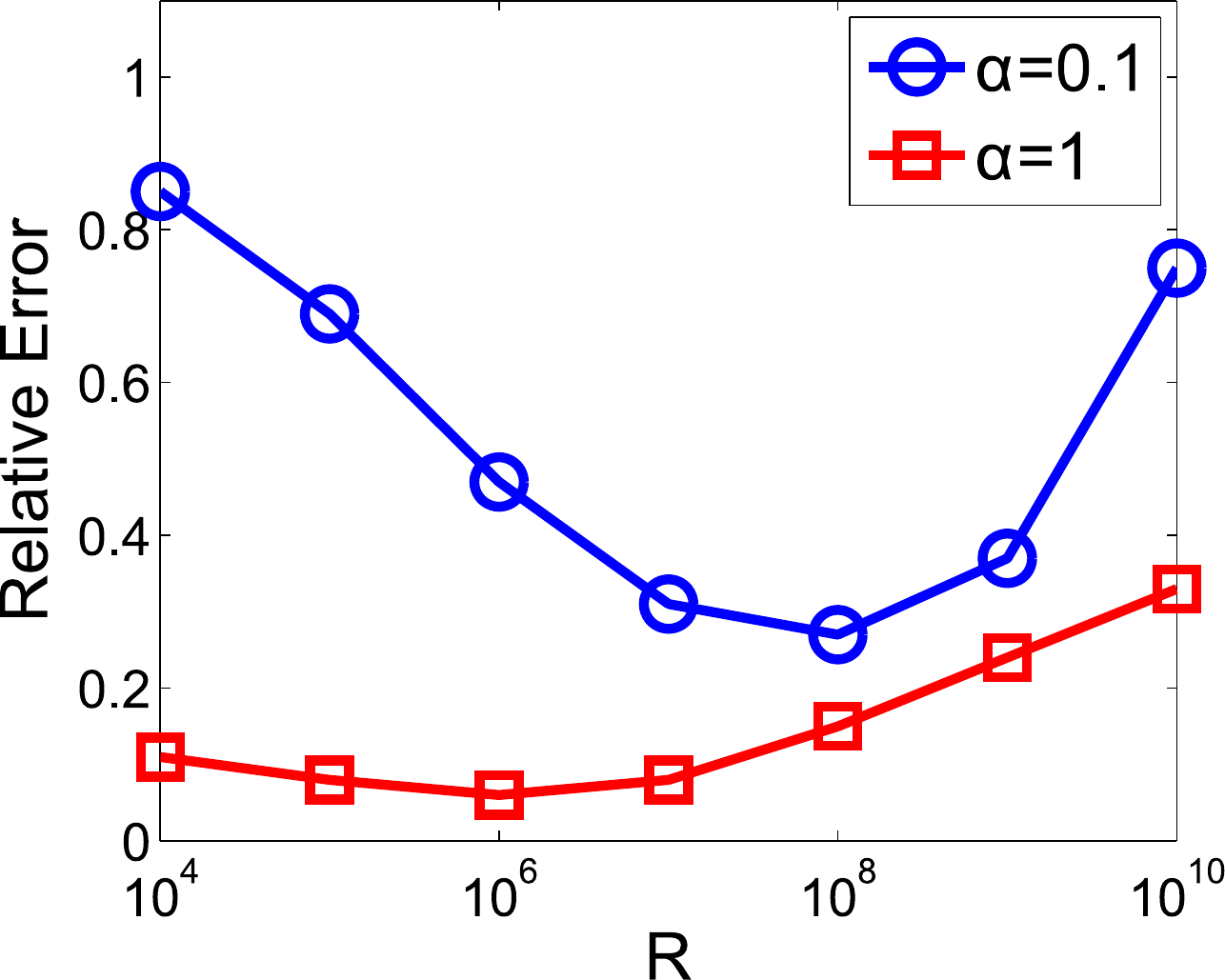}} 
  \caption{\small Choice of $R$ in the Kalman Filter} 
\label{fig:kalman_R}
\end{figure}

\vspace{0.05in}
\noindent {\bf Choice of $R$ in the Kalman Filter.}
Since the process noise $Q$ is intrinsic to the time-series data of interest,  it can be determined by domain users who have good understanding about the process to monitor or have access to some historic data.   What is not straight-forward is the choice of $R$, the Gaussian noise we propose to use in order to approximate the Laplace perturbation noise.  Figure~\ref{fig:kalman_R:a} plots the utility of the released time-series with respect to varying $R$ values when using first $10\%$ of the data series versus using the entire series.   Figure~\ref{fig:kalman_R:b} plots the utility with respect to varying $R$ values with $\alpha$ set to $0.1$ versus $1$.   As can be seen in Figure~\ref{fig:kalman_R:a}, when using $10\%$ of the data, the optimal $R$ value is $10^4$, as opposed to $10^6$ when using the entire series.  Similarly in Figure~\ref{fig:kalman_R:b}, when $\alpha = 1$, the optimal $R$ value is $10^6$, which is a hundred times less than $10^8$, the optimal $R$ for $\alpha = 1$.  Both these results confirm our finding in Theorem 3 that the optimal $R$ value is proportional to $T^2/\alpha^2$.

\begin{figure} [t]
  \centering 
  \subfigure[Relative Error]{ 
    \label{fig:particle_N:a} 
    \includegraphics[width=0.45\columnwidth]{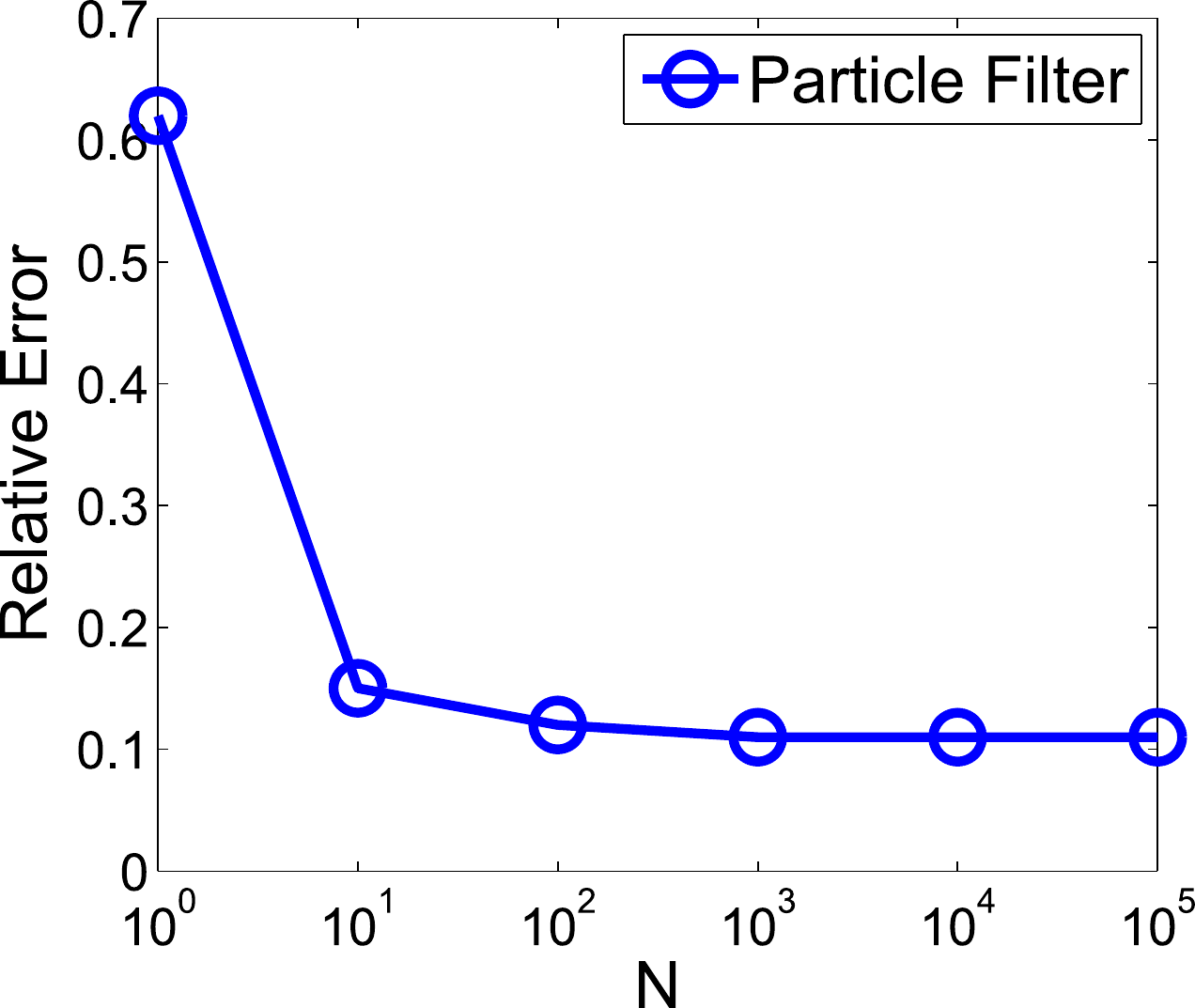}} 
  \hspace{0.02in} 
  \subfigure[Runtime]{ 
    \label{fig:particle_N:b} 
    \includegraphics[width=0.45\columnwidth]{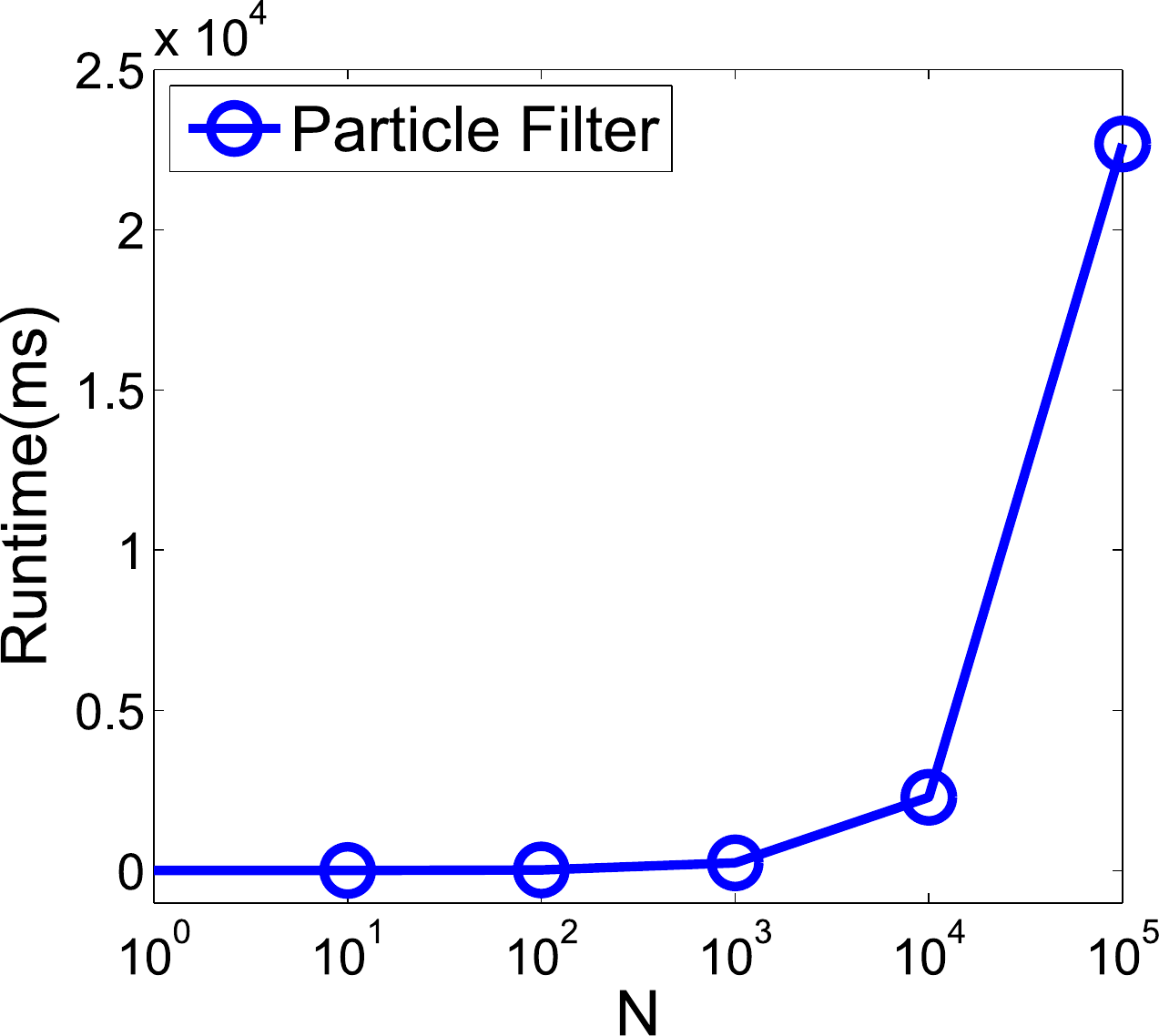}} 
  \caption{\small Choice of $N$ in Particle Filter} 
  \label{fig:particle_N} 
\end{figure}

\vspace{0.05in}
\noindent {\bf Choice of $N$ in Particle Filter.}
Due to the Monte Carlo nature of particle filtering methods, a larger number of particles implies better estimation of the posterior distribution.  However, $N$ cannot be infinitely large in the real-time scenario where fast response is crucial.  Here we examine the trade-off between accuracy and runtime of the particle filter.  Figure~\ref{fig:particle_N:a} plots the utility of particle filtering with different $N$ values.  As we expect, the relative error goes down as the number of particles increases and we observe no significant boost in accuracy when $N$ is greater than $10^3$.  On the other hand, a larger number of particles requires more processing time, as plotted  in Figure~\ref{fig:particle_N:b}.  Based on these results, we choose $N=10^3$ as the default value as it provides a good balance between accuracy and computation efficiency.

\begin{figure}
\centering
\includegraphics[width=0.95\columnwidth]{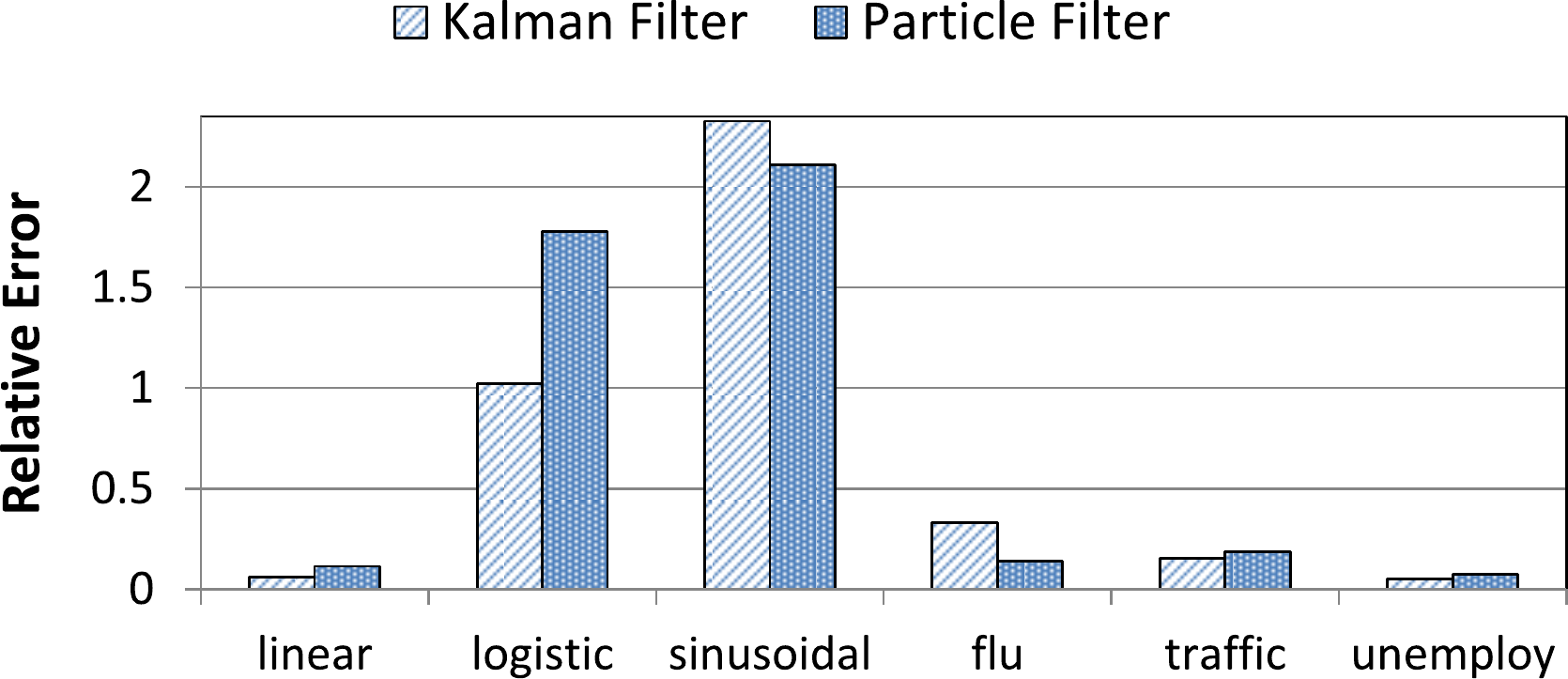}
\caption{ \small Comparison of Two Filtering Algorithms}
\label{fig:filter}
\end{figure}

\vspace{0.05in}
\noindent {\bf Kalman Filter vs. Particle Filter.} We compare two filtering methods across multiple data sets and summarize our findings in Figure~\ref{fig:filter}.   For \textit{logistic} and \textit{sinusoidal} data, both the Kalman filter and particle filter result in high relative errors due to the non-linearity in both data sets.  For the rest data sets, particle filter is comparable to the Kalman filter and even provides better utility for the \textit{flu} data set.   We conclude that particle filter is more robust since it relies on one parameter $N$ only and $N$ is independent of data sets, while the Kalman filter may provide optimal posterior estimation on condition that the Gaussian variance $R$ is wisely chosen.

\subsection{Effects of Sampling}
In the following studies, we apply sampling techniques on top of filtering and examine the advantage of adaptive sampling.

\begin{figure} 
  \centering 
  \subfigure[Choice of $\theta$]{ 
    \label{fig:adaSample:a} 
    \includegraphics[width=0.45\columnwidth]{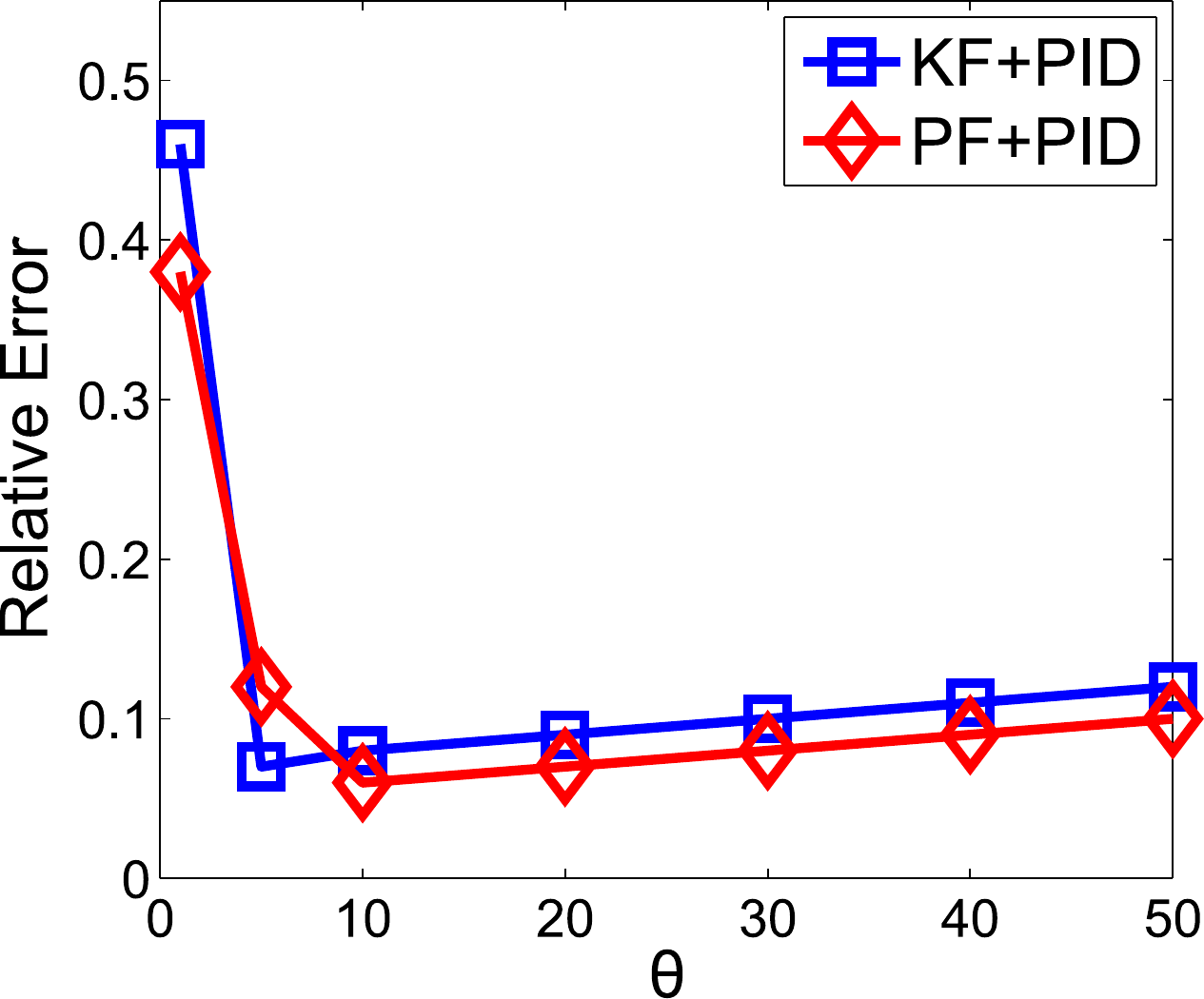}} 
  \hspace{0.02in} 
  \subfigure[Choice of $M$]{ 
    \label{fig:adaSample:b} 
    \includegraphics[width=0.45\columnwidth]{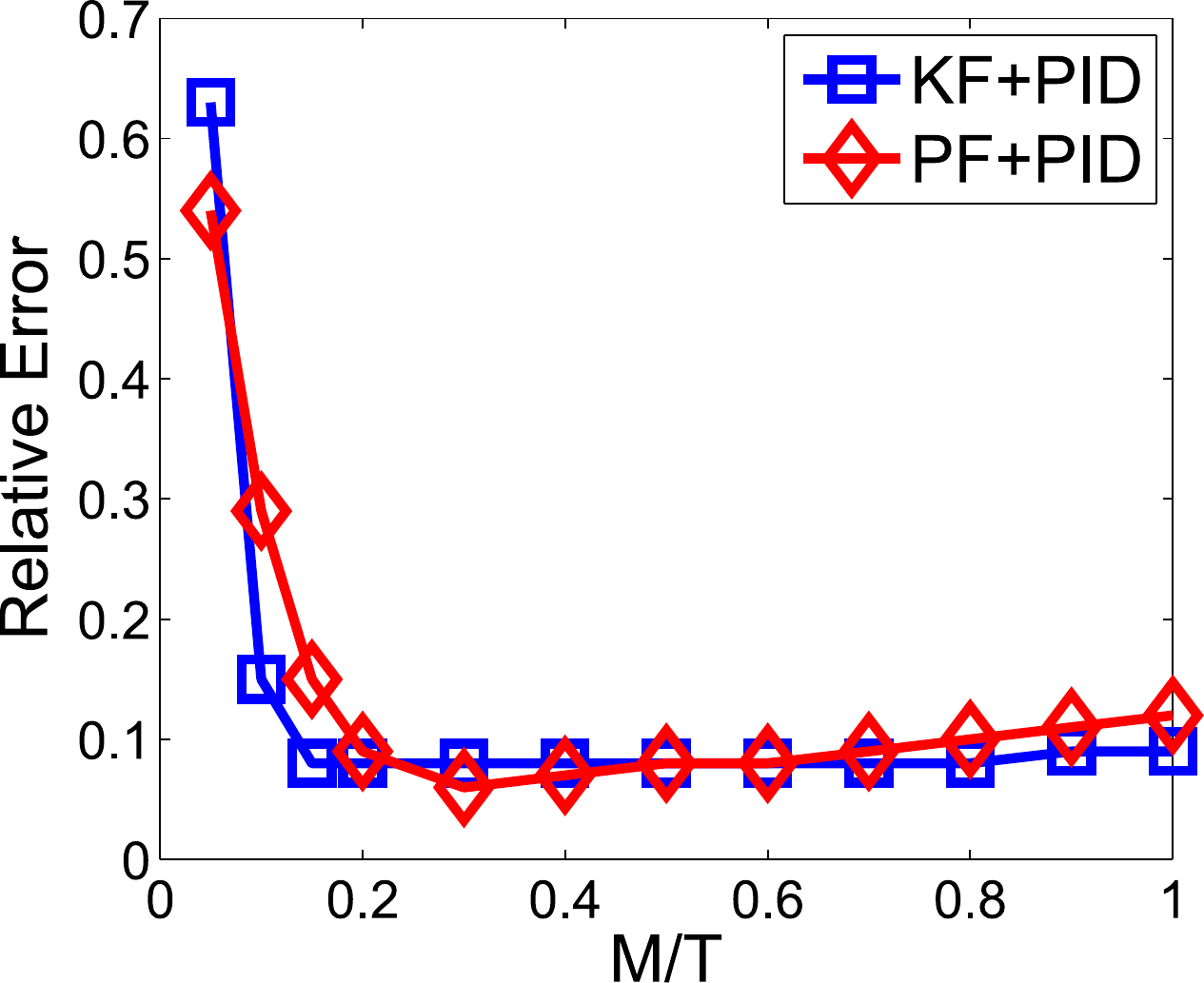}} 
  \caption{\small Choice of Adaptive Sampling Params} 
  \label{fig:adaSample} 
\end{figure}

\vspace{0.05in}
\noindent \textbf{Parameters for Adaptive Sampling.}  We first study the settings of adaptive sampling parameters $\theta$ and $M$.   Recall $\theta$ represents the magnitude of sampling interval adaptation and $M$ represents the maximum number of samples allowed for each application.  
Figure~\ref{fig:adaSample:a} shows the impact of $\theta$.  Both the Kalman filter and particle filter show similar utility results and trends as $\theta$ varies.  We observe that the error is prohibitive when $\theta=1$, due to insufficient interval adjustment. In that case, $I'$ is set to $1$ most of the time, according to Equation~(32), and the application quickly exhausts the privacy budget given.      The optimal $\theta$ value for the Kalman filtering method is $5$, while that of the particle filtering method is observed at $\theta=10$.  Both filtering methods result in slightly increased errors as $\theta$ increases beyond the optimal point, due to enlarged sampling interval and hence a higher prediction error between two adjacent samples.  However, the increase is insignificant compared to the extreme case where $\theta=1$.  Therefore, we conclude that FAST algorithms are robust to $\theta$ as we avoid apparent,  extremely small values.   Similarly, we state the same conclusion for the maximum number of samples $M$.   As shown in Figure~\ref{fig:adaSample:b}, we observe robust performance of FAST to the ratio $M/T$ as long as it is not deliberately set to be $M/T< 0.1$.  The optimal performance of FAST with the Kalman filter in this setting is achieved at $M=15\%T$ and with particle filter the optimal performance is at $M=25\%T$.  We record the findings above and use them for our other empirical studies.  

We also study the impact of the control gains $(C_p, C_i, C_d)$ as well as the integral time $T_i$.  We find that as long as the control gains are set according to the common practice: $proportional > integral > derivative$, the error variation between different settings is insignificant and is likely to be introduced by randomness.  Hence we omit the detailed results and conclude that there's no extra  ``rule of thumb''  beside the common practice for tuning the controller gains.  Similarly, as the integral time increases, the resulting error shows no clear trend.  We consider the above control parameters as non-influential in our system.

\begin{figure}
\centering
\includegraphics[width=0.5\columnwidth]{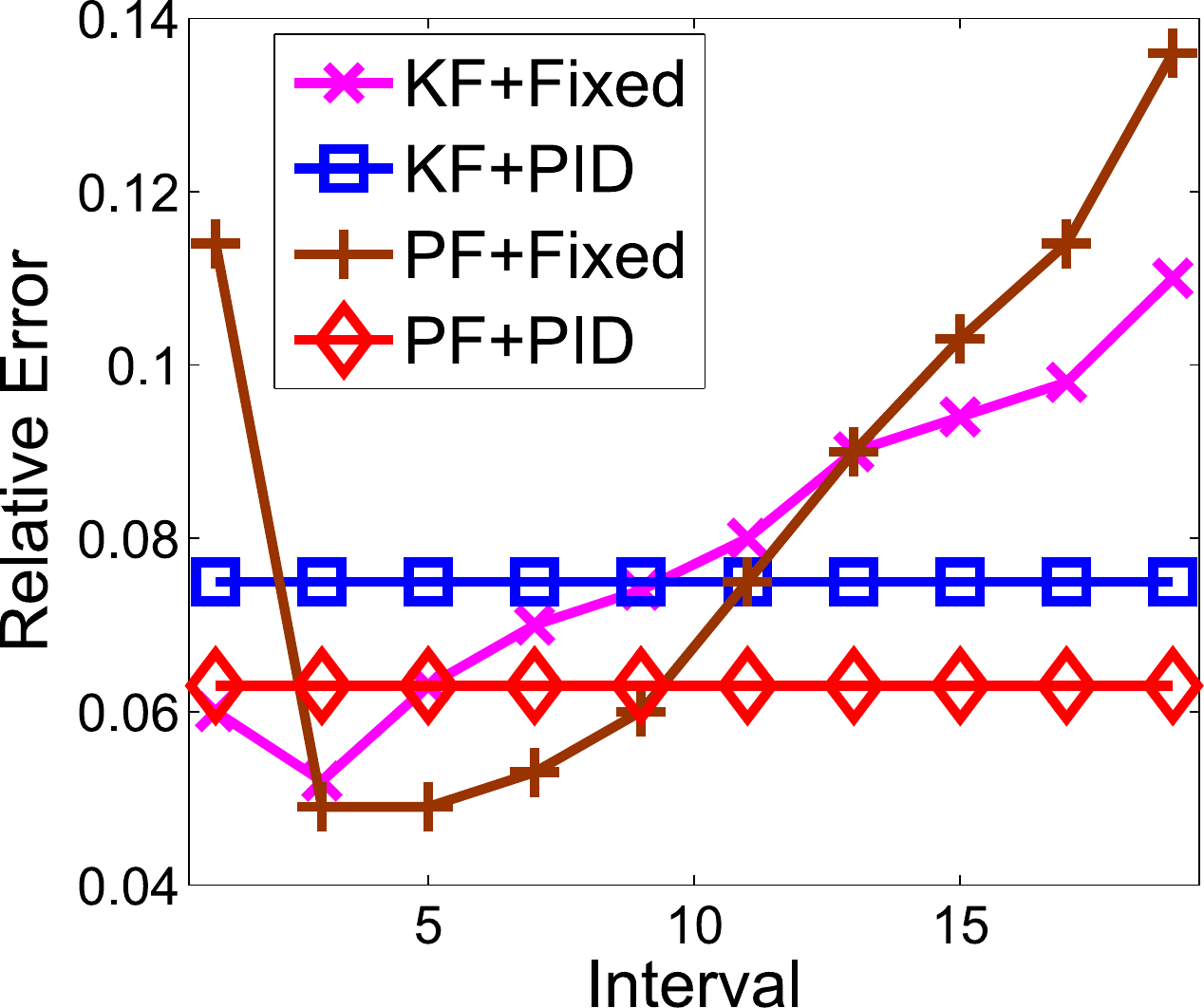}
\caption{ \small Fixed-Rate Sampling vs. Adaptive Sampling}
\label{fig:fixed}
\end{figure}

\vspace{0.05in}
\noindent \textbf{Fixed-Rate vs. Adaptive Sampling.}
We now compare the performance of adaptive sampling, denoted as PID in Figure~\ref{fig:fixed}, with fixed-rate sampling, while varying the sampling interval for fixed-rate algorithm from $1$ to $20$.  For our adaptive approach, we use the optimal $M$ setting for each filtering method.  However, there's a wide range of $M$ to choose from, which provides equivalent level of utility, according to our previous study.

The result is shown in Figure~\ref{fig:fixed}.   As the sampling interval increases, i.e. from 1 to 3, fixed-rate sampling shows reduced average relative error of various scales.  This phenomenon can be interpreted by the reduction of perturbation error resulting from less frequent queries. As the interval further increases, from 5 to 20, the error starts to rise, which can be explained by the accumulation of prediction error due to longer intervals between adjacent samples.  The optimal sampling interval, which is $3$ and $5$ for \textit{linear} data set, may not be known \textit{apriori} and may differ from dataset to dataset.  We found that the performance of adaptive sampling with no prior knowledge is comparable to the optimal fixed-rate despite the filtering method in use, which confirms once again the advantage of FAST adaptive framework.

\subsection{Utility vs. Privacy}

\begin{figure} 
  \centering 
  \subfigure[Overall Performance]{ 
    \label{fig:linear:a} 
    \includegraphics[width=0.47\columnwidth]{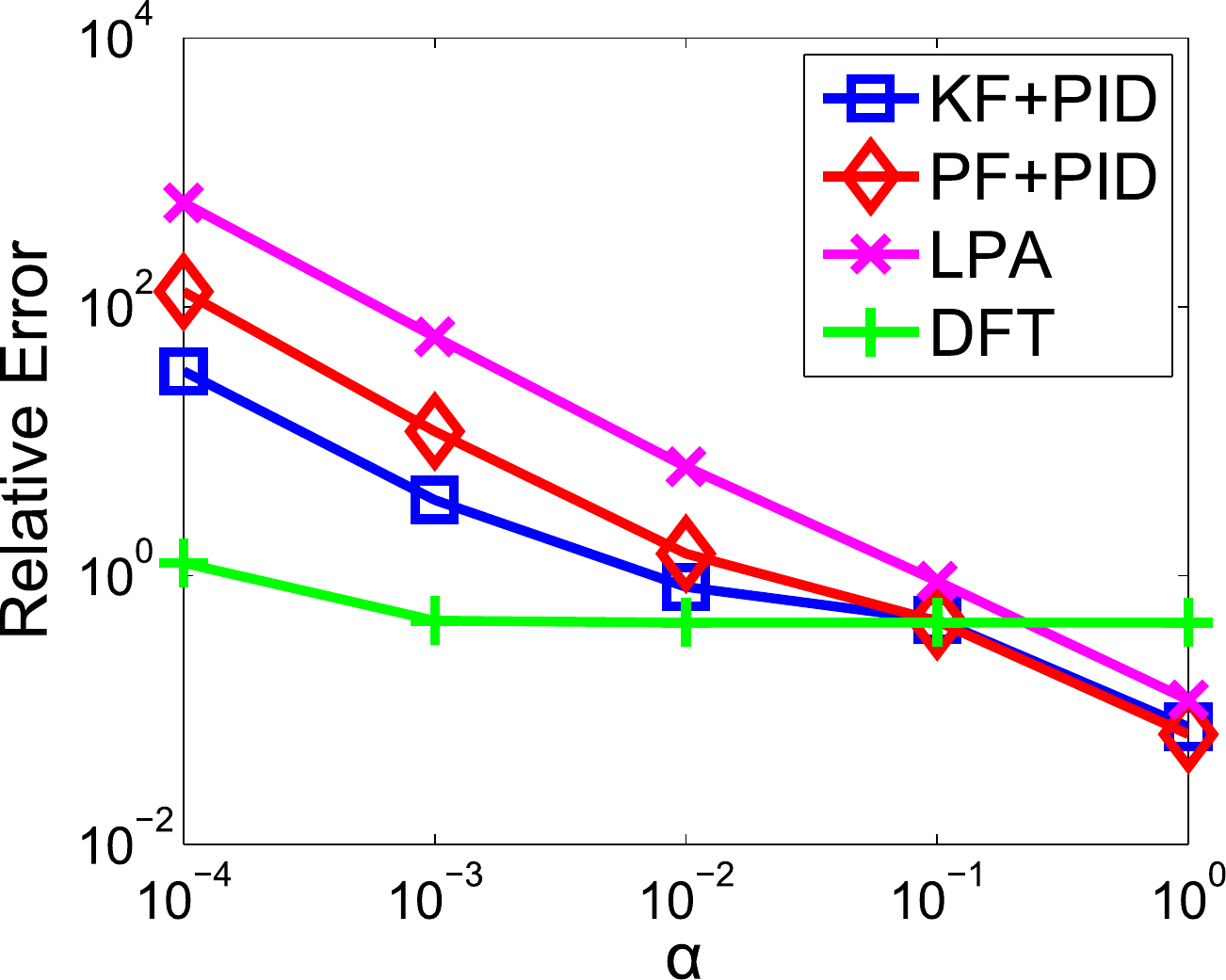}} 
  \hspace{0.02in} 
  \subfigure[Performance with Larger Budget]{ 
    \label{fig:linear:b} 
    \includegraphics[width=0.45\columnwidth]{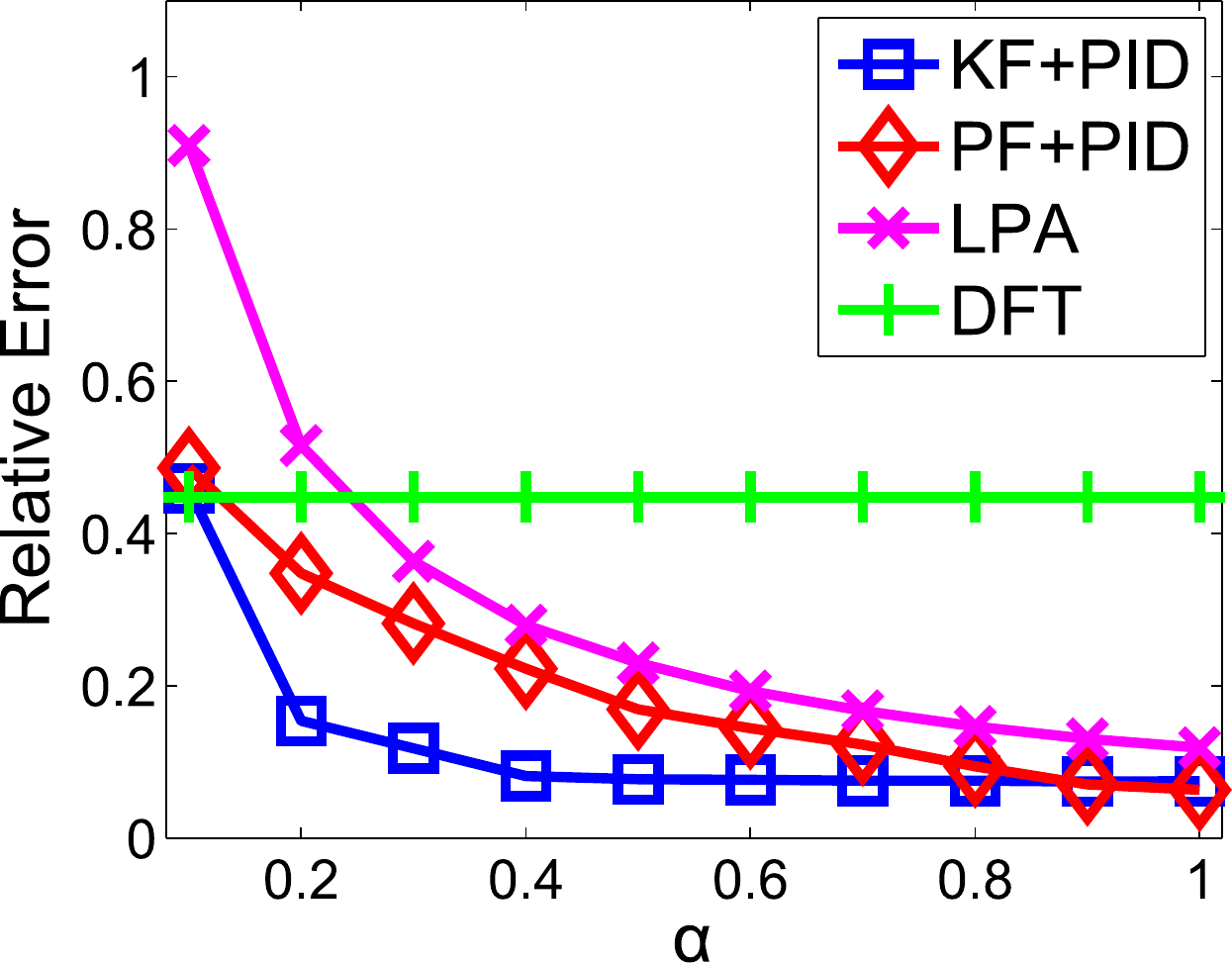}} 
  \caption{\small Utility vs. Privacy with Linear Data} 
  \label{fig:linear} 
\end{figure}

We examine the trade-off between utility and privacy in FAST, in comparison to the baseline LPA algorithm and the DFT algorithm. We note that DFT algorithm can be only applied offline and was run using the entire series in our experiments while FAST was run real-time.     Figure~\ref{fig:linear} shows the empirical study conducted with the \textit{linear} data set.   Figure~\ref{fig:linear:a} plots the relative error against a wide range of privacy budget values, from $10^{-4}$ to $1$.  As we relax the privacy level and increase the privacy budget $\alpha$, all four methods show reduced relative errors, to different extents. We observe that the DFT algorithm with off-line processing provides highest utility when $\alpha \in [10^{-4}, 10^{-1}]$.  However, no significant utility improvement can be seen when $\alpha \ge 10^-3$ due to its dominant reconstruction error.   On the other hand, FAST algorithms, i.e. {KF+PID} and {PF+PID}, consistently outperform LPA with reduced perturbation error.   When compared with DFT, FAST algorithms provide comparable utility and even outperform DFT when $ \alpha\in [10^{-1},1]$.  Figure~\ref{fig:linear:a} presents a closer look at this privacy budget interval, where FAST algorithms outperform both existing methods, providing high utility without compromising the privacy guarantee.

\begin{figure} 
  \centering 
  \subfigure[Logistic Data]{ 
    \label{fig:nonlinear:a} 
    \includegraphics[width=0.46\columnwidth]{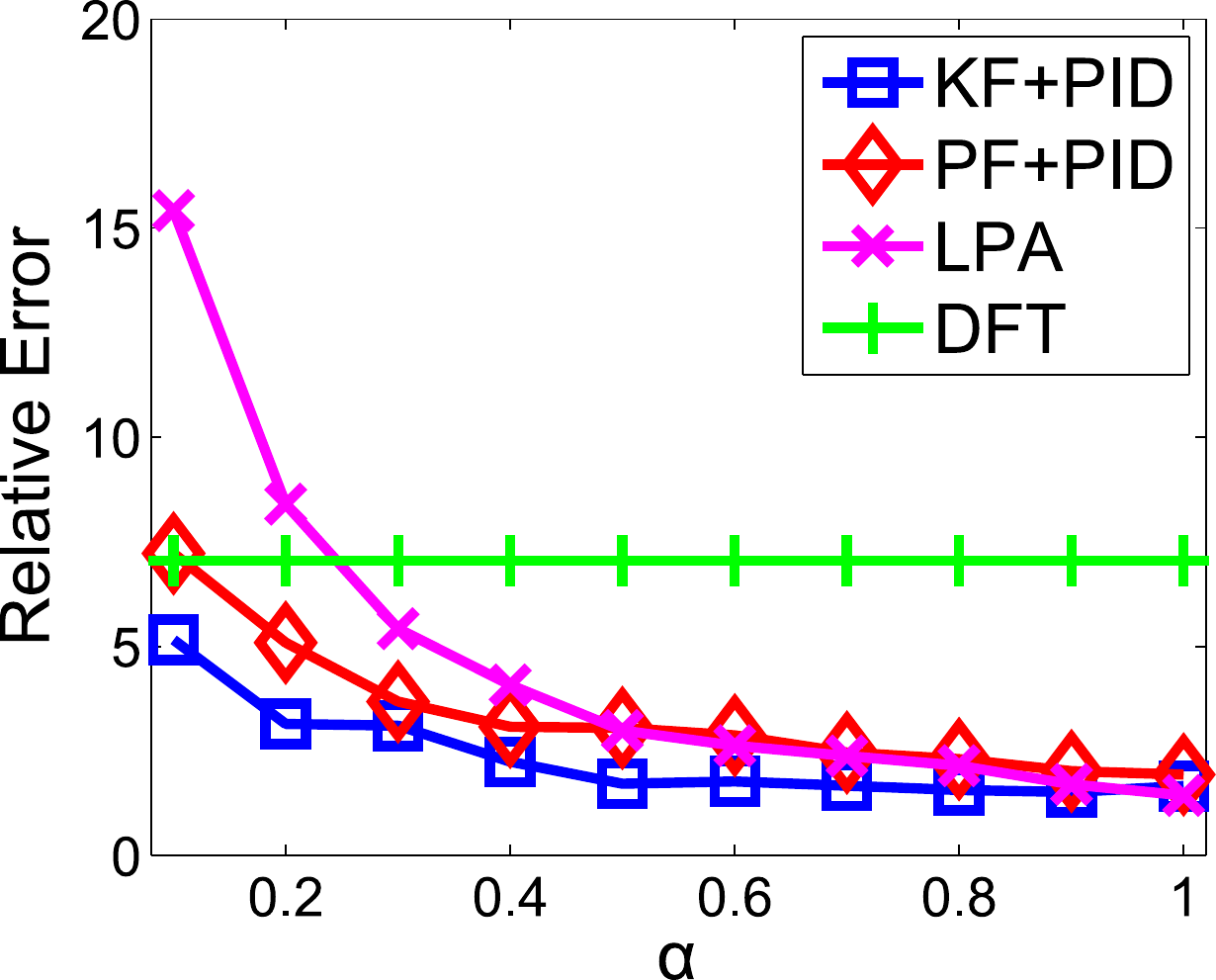}} 
  \hspace{0.02in} 
  \subfigure[Sinusoidal Data]{ 
    \label{fig:nonlinear:b} 
    \includegraphics[width=0.45\columnwidth]{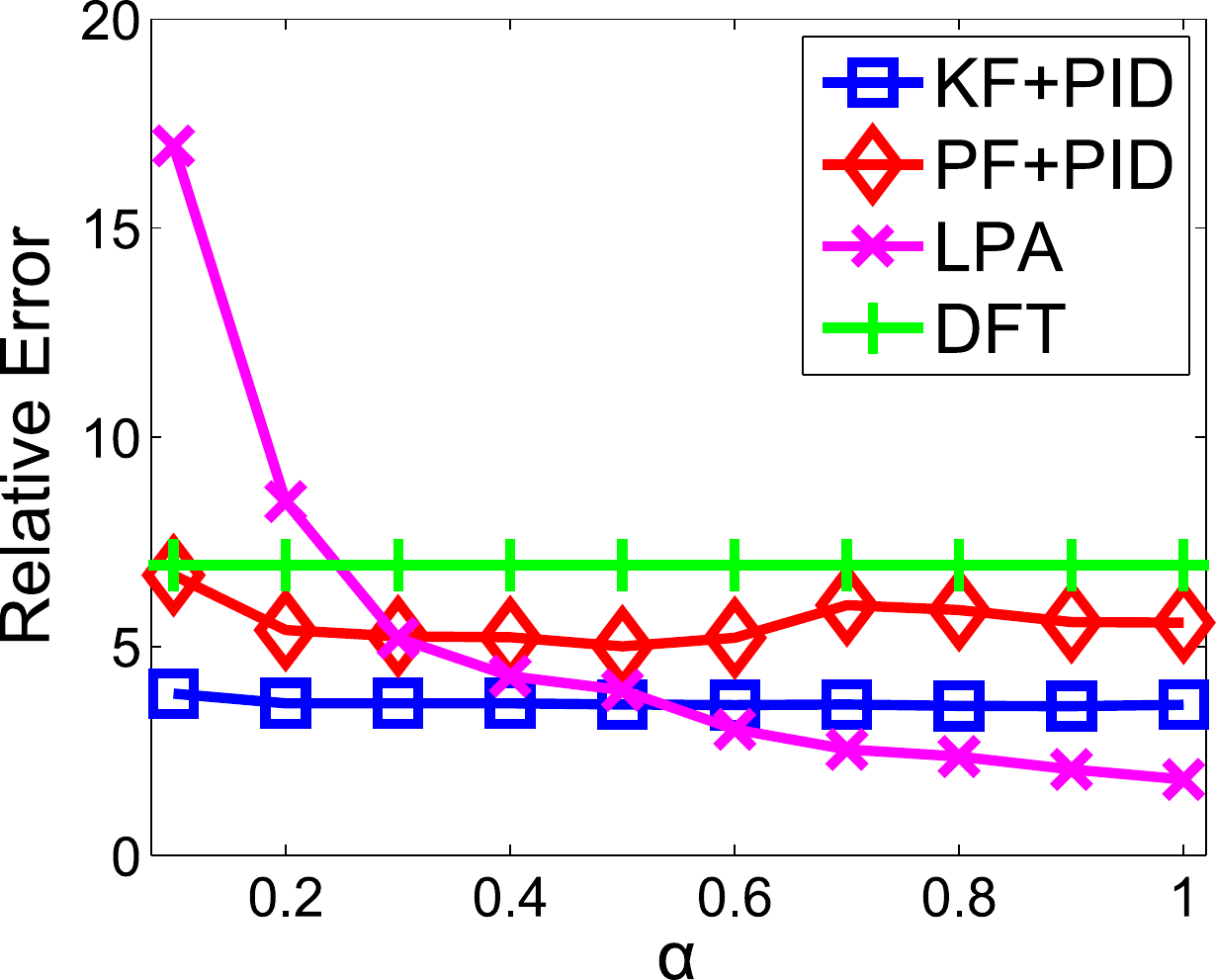}} 
  \caption{\small Utility vs. Privacy with Non-Linear Data} 
\vspace{-0.2cm}
  \label{fig:nonlinear} 
\end{figure}

In addition, we conducted the same trade-off studies with two synthetic data sets generated by non-linear models, in order to study the robustness of FAST framework when the internal state-space model cannot precisely capture the data dynamics.   Figure~\ref{fig:nonlinear} summarizes our findings.  For the \textit{logistic} data set, FAST algorithms still outperform two existing methods, even though the magnitude of error is higher.   For the \textit{sinusoidal} data set, LPA algorithm appears to be the best when $\alpha$ is between $0.5$ and $1$.  The reason is that FAST algorithms with linear state model and sampling may lose partially the periodic property of the sinusoidal data set.  We observed similar results from the study with real data sets, where the \textit{flu} data contains periodic peaks and the rest, i.e. \textit{traffic} and \textit{unemploy}, are nearly linear.

\subsection{Detection and Correlation}

In this study, we explore FAST performance with respect to utility metrics besides the standard relative error.   In particular, we studied the F1 metric in outbreak detection with the released series and also performed correlation analysis between the released and original aggregate series.

\begin{figure}
\centering
\includegraphics[width=1\columnwidth]{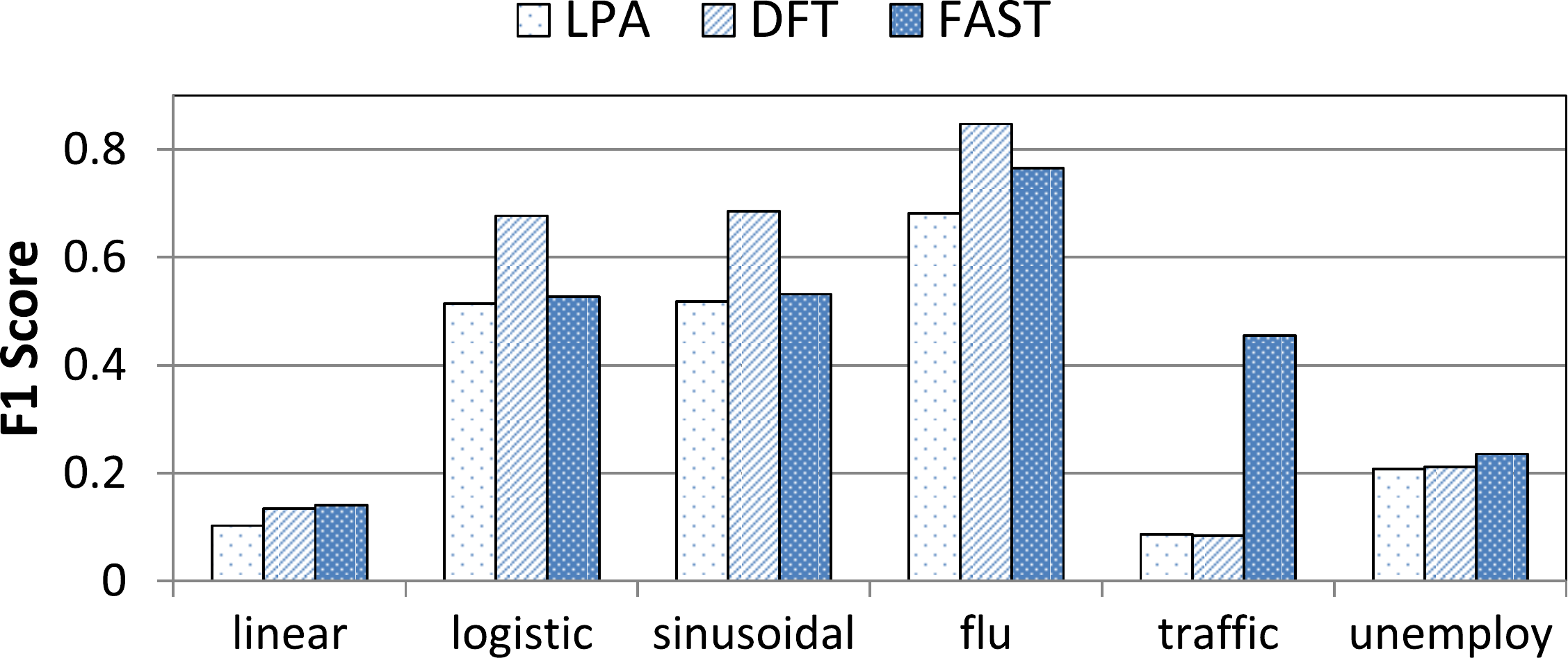}
\caption{ \small F1 Metric of Outbreak Detection}
\label{fig:f1}
\end{figure}

There have been extensive studies on effective methods in epidemic outbreak detection and various definitions of an outbreak/signal period have been adopted~\cite{Wagner01,Kleinman06, Pelecanos10}.  We take a simplified interpretation of outbreak similar to Pelecanos et al.~\cite{Pelecanos10} and define a target event/signal as a significant increase between two adjacent aggregate values. Usually the threshold of increase can be given by users according to the application.  In our empirical study,  we set this threshold to be $5\%$ of the median value in the original aggregate series, in order to mitigate the effects of extremely small or large values.  Figure~\ref{fig:f1} compares FAST solution against existing methods and shows the F1 metric of event/signal detection in the released series across multiple data sets.   We observe that FAST consistently outperforms LPA algorithm and provides comparable utility to DFT algorithm. In many cases, i.e. \textit{traffic} and \textit{unemploy} data sets, our approach provides even better utility than DFT.

\begin{figure}
\centering
\includegraphics[width=1\columnwidth]{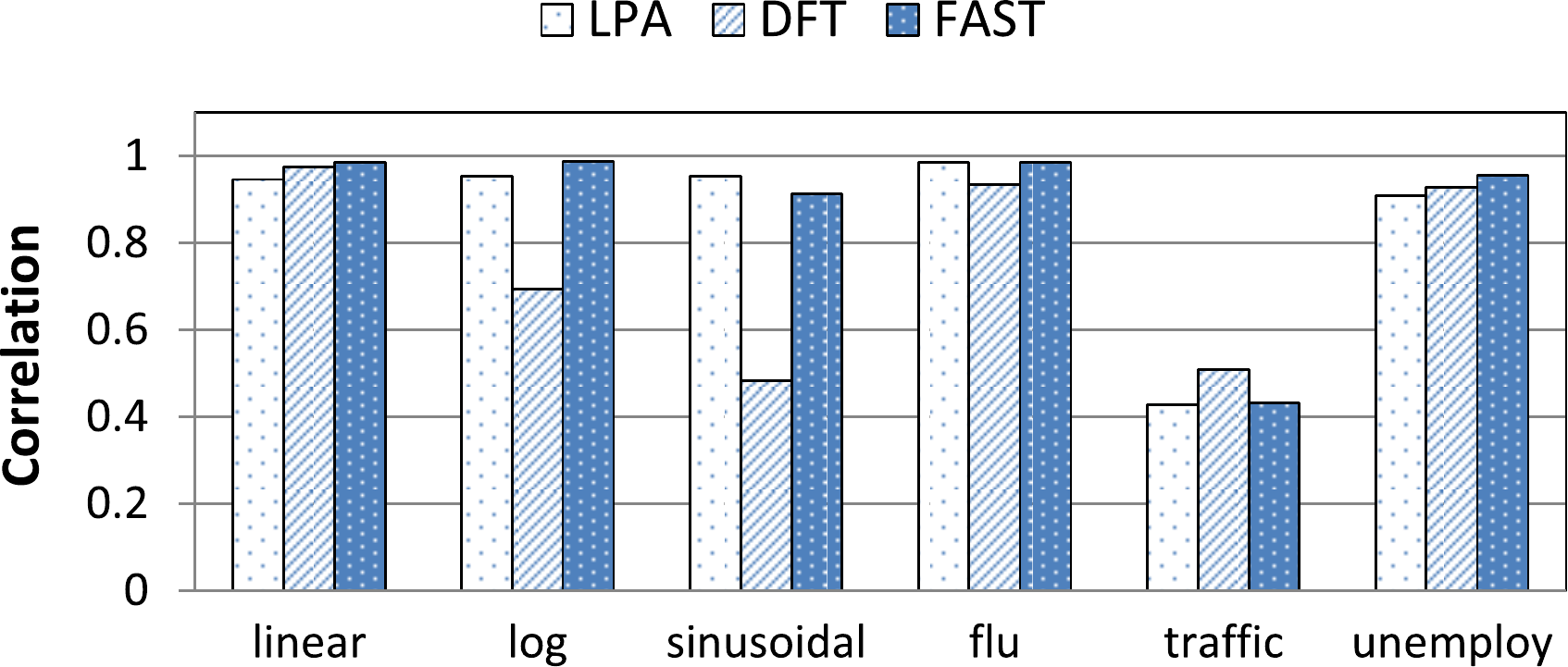}
\caption{ \small Spearman's Rank Correlation between Original Series and Released Series}
\label{fig:corr}
\end{figure}

As for correlation analysis, we measure the similarity between the private, released series and the original series with the Spearman's rank correlation.  Figure~\ref{fig:corr} summarizes the correlation study on FAST against existing methods.  We observe that the DFT algorithm doesn't preserve the ranking order in released series for \textit{sinusoidal} data.  In contrast, FAST algorithm provides robust performance across all data sets, even when DFT fails to produce correlated release.

To summarize the above evaluations, we find that FAST is comparable to the best existing method in each study while providing real-time releases, in contrast to DFT, which is only compatible in offline or batch mode.  In many cases, FAST outperforms both existing methods, e.g, \textit{traffic} data in outbreak detection.  Both evaluations confirm that FAST improves the utility of released data under differential privacy with filtering and adaptive sampling techniques, despite different data dynamics. We believe that FAST will enable a wider range of monitoring applications with the real-time feature and the adaptation strategies.

\vspace{-0.1cm}
\section{Conclusion}
We have proposed FAST, an adaptive framework with filtering and sampling  for monitoring real-time aggregate under differential privacy. The key innovation is that our approach utilizes feedback loops based on observed values to dynamically adjust the prediction/estimation model as well as the sampling rate.  To minimize the overall privacy cost, FAST uses the PID controller to adaptively sample long time-series according to detected data dynamics.  As to improve the accuracy of data release per time stamp, FAST uses filtering to predict data values at non-sampling points and to estimate true values from perturbed values at sampling points.  Our experiments with real-world and synthetic data sets show that it is beneficial to incorporate feedback into both the estimation model and the sampling process.  The results confirmed that our adaptive approach improves utility of time-series release and has excellent performance even under small privacy cost.  

As for the future, we plan to expand our solution to enable monitoring of differentially private spatial-temporal statistics, for example, real-time traffic conditions at all intersections of a city.

\appendices

\section{The Kalman Filter Posterior Analysis}
If the Kalman filter is applied every time stamp, which is equivalent to fixed-rate sampling with $I=1$, we derive the posterior error variance as follows:
\begin{align*}
var(\hat{x}_k - x_k) = E(\hat{x}_k - x_k)^2 - E^2(\hat{x}_k - x_k)
\end{align*}
Since $\omega$ and $\nu$ are white noise and mutually, serially independent, by definition of $\hat{x}_k$ we get
\begin{align*}
 E(\hat{x}_k - x_k) = 0.
\end{align*}
Therefore, we will only need to estimate the first term in Equation~(33).  Substituting Equation~(14) leads to
\begin{align*}
E(\hat{x}_k - x_k)^2 &= E[ (1-K_k)(\hat{x}_k^- - x_k) + K_k \nu]^2   \nonumber \\
&= E[ (1-K_k)(\hat{x}_{k-1} - x_k) + K_k \nu ]^2 \nonumber \\
&= E[ (1-K_k)(\hat{x}_{k-1} - x_{k-1})   \nonumber \\
& \qquad{} +  (1-K_k)(x_{k-1} - x_{k}) + K_k \nu ]^2 \nonumber \\
&= E(1-K_k)^2(\hat{x}_{k-1} - x_{k-1})^2  \nonumber \\
& \qquad{} + E[(1-K_k)^2\omega^2] + E (K_k^2 \nu^2)  \nonumber \\
&= (1-K_k)^2 E(\hat{x}_{k-1} - x_{k-1})^2  \nonumber \\
& \qquad{}+  (1-K_k)^2Q + K_k^2 \frac{2T^2}{\alpha ^ 2} 
\end{align*}
where $\alpha$ is the overall privacy budget and $T$ is the lifetime of the time-series.

Substituting the Kalman gain, i.e. Equation~(17), into the above variance, we get
\begin{align*}
E(\hat{x}_k - x_k)^2 = \frac{ R^2 [ E(\hat{x}_{k-1} - x_{k-1})^2  + Q]}{(P_k^- + R)^2} +\frac{2{P_k^- }^2 T^2}{(P_k^- + R)^2 \alpha^2}
\end{align*}
Apply gradient descendant method to minimize the current estimation error,  we obtain the following result:
\begin{align*}
R= \frac{T^2}{\alpha^2} \frac{2 P_k^- }{ E(\hat{x}_{k-1} - x_{k-1})^2  + Q}
\end{align*}

\section{Fixed-Rate Sampling Posterior Analysis}

\subsection{Posterior Error at a Sampling Point}
When sampling techniques are combined with the Kalman filter, we would not obtain a measurement at every time stamp except for the sampling points.  Assume we adopt a fixed rate sampling technique with interval $I$ and the current time stamp $k$ is a sampling point.  The prior estimate at time stamp $k$ is actually the posterior estimate of time stamp~$k-I$:
\begin{align*}
\hat{x}_k^- = \hat{x}_{k-I}
\end{align*}

By introducing dummy terms, we get
\begin{align*}
\hat{x}_k - x_k &= (1-K_k)(\hat{x}_{k-I} - x_k) + K_k \nu  \nonumber  \\
& =  (1-K_k)[ (\hat{x}_{k-I} - x_{k-I}) - (x_k - x_{k-I})] + K_k \nu \nonumber \\
& = (1-K_k)[ (\hat{x}_{k-I} - x_{k-I}) - \sum_{k-I+1}^{k}\omega_j ] + K_k \nu 
\end{align*}
According to the process model in Equation~(4),  $I$ independent white Gaussian process noise variables, i.e. $\omega_j$'s,  are introduced between time stamp $k-I+1$ and time stamp $k$.

Therefore,
\begin{align*}
E(\hat{x}_k - x_k ) &=(1-K_k)[E(\hat{x}_{k-I} - x_{k-I}) - \sum_{k-I+1}^{k} E\omega_j] + K_k E\nu \nonumber \\
&= 0
\end{align*}
and
\begin{align*}
var(\hat{x}_k - x_k ) &= E(\hat{x}_k - x_k )^2 \nonumber \\
&=(1-K_k)^2 E(\hat{x}_{k-I} - x_{k-I})^2  \nonumber \\
& \qquad{}+ (1-K_k)^2\sum_{k-I+1}^{k}E\omega_j^2 + K_k^2 E\nu^2 \nonumber \\
&= (1-K_k)^2 [E(\hat{x}_{k-I} - x_{k-I})^2 + IQ]  \nonumber \\
& \qquad{}+ 2K_k^2\frac{T^2}{I^2 \alpha^2}
\end{align*}
where $Q$ is the process noise covariance.

\subsection{Prediction Error at a Non-Sampling Point}
At non-sampling points, the prior estimate will be released and we will derive the error estimate below.  Assume the current time stamp $k$ is a non-sampling point and the most recent sample occurs at time stamp $k-t$. Therefore, by introducing dummy terms  and applying state model again, we get
\begin{align*}
\hat{x}_k^- -x_k &= \hat{x}_{k-t} - x_k \nonumber \\
&= \hat{x}_{k-t}  - x_{k-t} - (x_k - x_{k-t}) \nonumber \\
&=  \hat{x}_{k-t}  - x_{k-t} - \sum_{k-t+1}^{k}\omega_j
\end{align*}
and
\begin{align*}
E(\hat{x}_k^- -x_k ) = 0
\end{align*}
\begin{align*}
var(\hat{x}_k^- -x_k) &= E(\hat{x}_k^- -x_k )^2 \nonumber \\
&= E(\hat{x}_{k-t}  - x_{k-t})^2 +  \sum_{k-t+1}^{k}E\omega_j^2 \nonumber \\
&= E(\hat{x}_{k-t}  - x_{k-t})^2 + tQ
\end{align*}

\subsection{Overall Error}
Let $S$ denote the set of sampling points, for instance, $S = \{0, I, 2I, ...\}$.  Suppose $k \in S$ and let $var_j$ denote the error variance at any time stamp $j$.  The sum of error between time stamp $k$ and $k+I-1$, right before the next sample, can be found as follows
\begin{align*}
\sum_{k}^{k+I-1} var_j  
&= \sum_{k}^{k+I-1} [var_k + (j-k)Q] \nonumber \\
&= I \cdot var_k +  \sum_{0}^{I-1} jQ \nonumber \\
&= I \cdot var_k + \frac{I(I-1)}{2}Q
\end{align*}

The overall error for the entire time series can be found:
\begin{align*}
sumErr &= \sum_{k \in S} \sum_{k}^{k+I-1} var_j \nonumber \\
&= I \cdot \sum_{k \in S} var_k + |S|  \frac{I(I-1)}{2}Q \nonumber \\
&=  I \cdot \sum_{k \in S} var_k + \frac{T(I-1)}{2}Q
\end{align*}
since the size of $S$ is $T/I$.

We can further substitute each $var_k$ at sampling points. Therefore the sum of error variance over $T$ can be viewed as a function of interval length $I$
\begin{align*}
sumErr &=I^2 \cdot Q\sum_{k \in S}(1-K_k)^2  \nonumber \\
& \qquad {} + I \cdot [\sum_{k \in S}var_{k-I}(1-K_k)^2 +\frac{TQ}{2}] \nonumber \\
& \qquad {} + I^{-1} \cdot  \frac{2T^2}{\alpha^2} \sum_{k \in S}K_k^2 - \frac{TQ}{2}
\end{align*}
which is very challenging to minize \textit{a priori} due to the recursive filtering procedures.

\vspace{-0.1cm}
\section*{Acknowledgment}
This research was supported in part by NSF grant CNS-1117763, AFOSR grant \#12RSE136, and an Emory URC grant.

\bibliographystyle{IEEEtran}
\bibliography{reference}

\end{document}